\documentclass[11pt,a4paper,oneside,openany]{report}
\usepackage{nf-style}
\newcommand{\nfVersion}{0.107.0}
\newcommand{\nfTopCmds}{150}
\newcommand{\nfCmdPaths}{374}

\newcommand{\nfExportCmds}{31}
\newcommand{\nfPackages}{84}
\newcommand{\nfPyKLOC}{260}
\newcommand{\nfTestFiles}{1,239}
\newcommand{\nfTestFuncs}{16,721}
\newcommand{\nfADRs}{299}
\newcommand{\nfADRMax}{317}
\newcommand{\nfDashTabs}{30}
\newcommand{\nfDashGroups}{7}
\newcommand{\nfServeRoutes}{214}
\newcommand{\nfServerRoutes}{79}
\newcommand{\nfAuthzRoutes}{225}
\newcommand{\nfRunners}{6}
\newcommand{\nfAdapters}{11}
\newcommand{\nfAuditProfiles}{6}
\newcommand{\nfSLOs}{21}
\newcommand{\nfSLOGated}{10}
\newcommand{\nfSLOMeasured}{8}
\newcommand{\nfSLOTarget}{3}
\newcommand{\nfWorkflows}{20}
\newcommand{\nfDiffClaims}{10}
\newcommand{\nfDiffPass}{10}

\hypersetup{
  pdftitle={NovaFabric Technical Overview: The Black-Box Recorder for AI Agents (Second Edition)},
  pdfauthor={Mohsen Seyedkazemi Ardebili},
  pdfsubject={The black-box recorder for AI agents: architecture, capabilities, evidence},
  pdfkeywords={NovaFabric, AI agents, provenance, replay, tamper evidence, DSSE, RFC 3161, lineage, EU AI Act}}

\newcommand{\cliitem}[2]{\mbox{\texttt{\small #1}\ifnum#2>0\,{\color{nfViolet}\scriptsize\textsuperscript{#2}}\fi}\quad }

\begin{document}
\begin{titlepage}
\thispagestyle{empty}
\begin{tikzpicture}[remember picture,overlay]
  \fill[nfCanvas] (current page.south west) rectangle (current page.north east);
  \begin{scope}[opacity=0.06]
    \foreach \x in {0,8,...,210} \draw[white,line width=0.3pt]
      ([xshift=\x mm]current page.south west) -- ([xshift=\x mm]current page.north west);
    \foreach \y in {0,8,...,297} \draw[white,line width=0.3pt]
      ([yshift=\y mm]current page.south west) -- ([yshift=\y mm]current page.south east);
  \end{scope}
  \node[anchor=north west,text=nfLime,font=\sffamily\fontseries{mb}\footnotesize]
    at ([xshift=24mm,yshift=-24mm]current page.north west)
    {\addfontfeatures{LetterSpace=14}TECHNICAL OVERVIEW \quad\textperiodcentered\quad SECOND EDITION};
  \nfmarkat{([xshift=24mm,yshift=-56mm]current page.north west)}{18mm}
  \node[anchor=north west,text=nfOnDark,align=left]
    at ([xshift=23mm,yshift=-88mm]current page.north west)
    {\displayfont\fontseries{eb}\fontsize{60}{60}\selectfont NovaFabric};
  \node[anchor=north west,text=nfOnDark,align=left,text width=150mm]
    at ([xshift=24mm,yshift=-114mm]current page.north west)
    {\serifdisplay\itshape\fontsize{21}{27}\selectfont The black-box recorder for AI agents:\\
     sealed, replayable, diffable, auditable runs};
  \begin{scope}[shift={([xshift=37mm,yshift=-170mm]current page.north west)}]
    \foreach \v [count=\i from 0] in {Capture,Seal,Replay,Diff,Audit} {
      \node[draw=nfLime,line width=0.9pt,rounded corners=4.5mm,minimum height=9mm,
            minimum width=26mm,fill=nfRaised,
            text=nfOnDark,font=\sffamily\fontseries{mb}\small]
        (v\i) at (\i*32mm,0) {\v};
    }
    \foreach \i [evaluate=\i as \j using int(\i+1)] in {0,...,3}
      \draw[-{Latex[length=2.2mm]},nfLime,line width=0.9pt] (v\i.east) -- (v\j.west);
  \end{scope}
  \draw[nfOnDarkMuted!50!nfCanvas,line width=0.4pt]
    ([xshift=24mm,yshift=46mm]current page.south west) -- ([xshift=-24mm,yshift=46mm]current page.south east);
  \node[anchor=north west,text=nfOnDark,align=left,font=\sffamily\small]
    at ([xshift=24mm,yshift=42mm]current page.south west)
    {\bfseries Mohsen Seyedkazemi Ardebili\\[1mm]
     {\color{nfOnDarkMuted}\footnotesize mskazemi.github.io \textperiodcentered{} ORCID 0000-0002-1166-6559}};
  \node[anchor=north east,text=nfOnDark,align=right,font=\sffamily\small]
    at ([xshift=-24mm,yshift=42mm]current page.south east)
    {\bfseries Built from NovaFabric v\nfVersion\\[1mm]
     {\color{nfOnDarkMuted}\footnotesize October 2026 \textperiodcentered{} supersedes Edition 1 (v0.45.1, June 2026)}};
\end{tikzpicture}
\end{titlepage}

\pagenumbering{roman}
\thispagestyle{plain}
\vspace*{4mm}
{\sffamily\footnotesize\color{nfViolet}\addfontfeatures{LetterSpace=10}ABSTRACT}\par\vspace{4mm}
{\serifdisplay\fontsize{19}{25}\selectfont\color{nfNavy}NovaFabric: A Black-Box Recorder for AI Agents with Sealed, Replayable, Diffable and Auditable Runs\par}
\vspace{2mm}
{\sffamily\small Mohsen Seyedkazemi Ardebili}\par\vspace{7mm}
An AI agent run is no longer a function call. It is a cascade of prompts, model calls, tool
invocations, file and network effects and sub-agents. When such a run misbehaves, or an auditor asks what happened, logs and traces answer
poorly: they are mutable, partial and cannot be re-executed. This technical report describes
NovaFabric, an open-source (Apache-2.0) black-box recorder for AI agents that turns a run into
a sealed, replayable, diffable and auditable evidence capsule. The design rests on two formats
only. The Run Capsule is a directory of flat files and the source of truth; every database is a
rebuildable index over capsules. The Evidence Bundle is the signed export for an auditor.
Everything else follows existing standards: OpenTelemetry GenAI, OpenLineage, W3C PROV,
in-toto/DSSE, RFC 3161 and CycloneDX. A single-node run is the smallest case of a distributed one;
the schema supports a capsule hierarchy for Slurm arrays and Kubernetes jobs, though automatic
parent finalization is library-only today. The report covers the architecture and the engines behind the verb chain capture, seal, replay,
diff and audit: fail-closed mocked replay; a diff gate with a stable exit-code contract; DSSE
signatures that bind their intent and a capsule root that commits to file names; a capsule
treated as untrusted input; signer identity-trust levels with offline TSA and CRL checking;
maker-checker promotion against an operator-held trust list; and a deny-by-default authorization table for every dashboard route. It presents no
new experiments; the evaluation is in arXiv:2609.12582. Every count except the private ADR tally is generated by a probe script run on the public
source at the v\nfVersion{} tag (\nfTopCmds{} top-level commands, \nfPackages{} Python subpackages,
\nfTestFuncs{} test functions), every capability carries an explicit maturity label, and
honest-limits boxes state where the system stops.

\vspace{4mm}
{\sffamily\footnotesize\color{nfSlate}\textbf{Keywords:} AI agents, provenance, deterministic replay, tamper evidence, audit,
DSSE, RFC 3161, lineage, EU AI Act}
\clearpage

\thispagestyle{plain}
\vspace*{4mm}
{\sffamily\footnotesize\color{nfViolet}\addfontfeatures{LetterSpace=10}ABOUT THIS EDITION}\par\vspace{3mm}
{\serifdisplay\fontsize{19}{25}\selectfont\color{nfNavy}A fresh reading of the system, checked against its own source tree.\par}
\vspace{5mm}

\begin{minipage}[t]{0.58\linewidth}\vspace{0pt}
This is the second edition of the NovaFabric Technical Overview. The first edition was
written against v0.45.1 in June 2026. Since then the CHANGELOG has recorded further minor
release lines, from v0.46 through v\nfVersion. The CLI has grown from 63 commands to \nfTopCmds{} top-level commands. The
replay engine has a fail-closed contract, the dashboard has an authorization model, and
NovaSeal has identity-trust levels. Signatures, Evidence Bundles and maker-checker approvals
are verified against what they claim, and the dashboard and HTTP API run the same verifier
as the CLI. A capsule is treated as untrusted input once the workload has run. A
secure-by-default deployment chart now exists. This
edition re-derives the whole picture from the code at \textbf{v\nfVersion}.

Three rules governed the rewrite:
\begin{enumerate}
  \item \textbf{Numbers are generated, not typed.} Every count in this document, except the ADR count (the decision records are private), comes from
        a probe script (\fpath{verify\_facts.py}) run against the source tree. It
        introspects the CLI, builds the HTTP apps, counts ADRs and tests, and reads the SLO
        catalogue. Appendix~\ref{app:method} lists the probes.
  \item \textbf{Every capability carries a maturity label.} The project uses four labels:
        \works, \exper, \planned{} and \fdesign. Where prose and code disagreed, the code won,
        and the disagreement is noted.
  \item \textbf{Every diagram is rendered and inspected.} Every figure is produced from
        editable Mermaid source with one shared theme. Each was checked by eye against the
        facts it depicts.
\end{enumerate}
\end{minipage}\hfill
\begin{minipage}[t]{0.36\linewidth}\vspace{0pt}
\begin{tcolorbox}[enhanced,boxrule=0pt,colback=nfMist,arc=1.6mm,left=4mm,right=4mm,top=3mm,bottom=3mm,fontupper=\sffamily\footnotesize\RaggedRight]
{\bfseries\color{nfNavy}Editions}\par\vspace{1.5mm}
{\color{nfViolet}\bfseries Second Edition} \textperiodcentered{} October 2026\\
built from v\nfVersion{} (the release tag; every generated count was read from that tag, see Appendix~\ref{app:method})\par\vspace{2mm}
{\color{nfSlate}\bfseries Edition 1} \textperiodcentered{} June 2026\\
built from v0.45.1; kept unchanged alongside this edition for comparison\par\vspace{3mm}
{\bfseries\color{nfNavy}Author}\par\vspace{1mm}
Mohsen Seyedkazemi Ardebili\\
{\color{nfViolet}ORCID 0000-0002-1166-6559}\\
{\color{nfViolet}mskazemi.github.io}\\
{\color{nfViolet}github.com/MSKazemi/novafabric}\par\vspace{3mm}
{\bfseries\color{nfNavy}Licence of the software}\par\vspace{1mm}
Apache-2.0 \textperiodcentered{} Python ≥ 3.12
\par\vspace{3mm}{\bfseries\color{nfNavy}Copyright}\par\vspace{1mm}
\textcopyright{} 2026 Mohsen Seyedkazemi Ardebili
\end{tcolorbox}
\end{minipage}

\clearpage
\thispagestyle{plain}
\vspace*{4mm}
{\sffamily\footnotesize\color{nfViolet}\addfontfeatures{LetterSpace=10}AT A GLANCE \textperiodcentered{} v\nfVersion}\par\vspace{3mm}
\noindent\stat{\nfTopCmds}{top-level \cmd{nova} commands, \nfCmdPaths{} runnable command paths}\hfill
\stat{\nfPackages}{Python subpackages, ≈\nfPyKLOC{}K lines of source}\hfill
\stat{\nfTestFuncs}{test functions in \nfTestFiles{} test files; 90\,\% coverage floor}\hfill
\stat{\nfADRs}{Architecture Decision Records (ADR-0001 to ADR-0\nfADRMax), kept privately}
\par\vspace{6mm}
\noindent\stat{\nfServeRoutes}{HTTP routes on \cmd{nova serve}, each classified in a deny-by-default scope table}\hfill
\stat{\nfDashTabs}{dashboard tabs in \nfDashGroups{} journeys}\hfill
\stat{\nfAdapters}{framework adapters; \nfRunners{} execution runners}\hfill
\stat{\nfExportCmds}{\cmd{nova export-*} evidence exporters; \nfAuditProfiles{} audit profiles}
\par\vspace{12mm}
{\sffamily\footnotesize\color{nfViolet}\addfontfeatures{LetterSpace=10}THE SHAPE OF THE SYSTEM}\par\vspace{3mm}
\begin{nftable}
\begin{tabularx}{\linewidth}{@{}P{0.2\linewidth}L@{}}
\toprule
\thead{Layer} & \thead{What it is, in one line} \\
\midrule
Record & the Run Capsule (flat files, source of truth) and the Evidence Bundle (signed in-toto ZIP) \\
Capture & SDK and wire hooks, \nfAdapters{} framework adapters, MCP and API proxies, \nfRunners{} runners; fail-open \\
Trust & NovaSeal: DSSE signature, RFC~3161 timestamp, append-only Merkle log, identity-trust level \\
Analyse & five replay modes with a fail-closed contract; structural diff with an exit-code contract; lineage \\
Govern & OPA/Rego gates, maker-checker, hash-chained audit log, \nfExportCmds{} exporters, erasure receipts \\
Serve \& scale & deny-by-default dashboard API, multi-user server, Go collector, object store as source of truth \\
\bottomrule
\end{tabularx}
\end{nftable}
\vspace{6mm}
{\sffamily\footnotesize\color{nfSlate}Maturity labels used throughout:\quad \works\quad \exper\quad \planned\quad \fdesign}
\clearpage

\tableofcontents
\clearpage
\pagenumbering{arabic}

\chapter{Executive summary}\label{ch:summary}

\begin{pullquote}NovaFabric is the black-box recorder for AI agents. It turns every important
agent run into a sealed, replayable, diffable and auditable evidence capsule.\end{pullquote}

An AI agent run is no longer a function call. It is a cascade of prompts, model calls,
tool invocations, file and network effects and sub-agents, often spread across a laptop,
a cluster and several providers. When such a run misbehaves, or an auditor asks what
happened, the questions that matter are concrete. What exactly crossed the wire? Can the
record be trusted? Can it be inspected or re-run months later? What changed between two
runs, and what else did it affect? NovaFabric answers these questions with one product
primitive and one verb chain (Figure~\ref{fig:fig-01-verb-chain}).

\nffigure{fig-01-verb-chain}{The product verb chain. Each verb is one \cmd{nova} command
and leaves one durable artefact behind.}

\paragraph{One record format, from laptop to cluster.} The \emph{Run Capsule} is a
directory of flat files. It is the source of truth, and every database in the system is
a derived index that can be rebuilt from capsules. The \emph{Evidence Bundle} is the
signed export handed to an auditor. These are the only two top-level formats the project
invents; everything else aligns with existing standards: OpenTelemetry GenAI~\cite{otel-genai},
OpenLineage~\cite{openlineage}, W3C PROV~\cite{provdm}, in-toto/DSSE~\cite{intoto,dsse}, RFC~3161~\cite{rfc3161} and CycloneDX~\cite{cyclonedx}. A single-node run is
treated as the smallest case of a distributed run. The same envelope and the same
capsule hierarchy are designed to carry a Slurm array or a Kubernetes job; automatic parent
finalization is library-only today (Chapter~\ref{ch:scale}). The system and an evaluation of it
were first reported in~\cite{nf-paper}; this report describes the current release in full.

\paragraph{Contributions of this report.} This is a reference report, not a new measurement
study. It contributes: (i) one account of the architecture of the current release in which
each statement about behaviour was checked against the code at the release tag
(Appendix~\ref{app:method}); (ii) the contracts of the main engines stated as testable
guarantees together with their limits, namely what each replay mode re-executes and refuses,
the exit-code contract of the diff gate, the verdict a sealed capsule produces, and the two-person
control paths; (iii) a trust model that separates what a signature, a timestamp and a log entry
each prove from what they do not, including the cases in which a verifier reports a capsule
as intact but not trusted; (iv) a catalogue of limits, with a maturity label on every
capability and honest-limits boxes, including the defects found while reading the text against
the code; and (v) a method for keeping a technical report true to a moving code base: counts
generated by probes, an executable differentiation table (Section~\ref{sec:difftable}) and a public script that re-derives the counts
(Appendix~\ref{app:method}).

\paragraph{Relation to arXiv:2609.12582.} The companion paper~\cite{nf-paper} presents the
system at an earlier release together with an empirical evaluation (research questions RQ1 to
RQ8). This report presents no new experimental campaign, and it neither restates nor extends
those measurements. It describes engine changes made since that evaluation, among them a seal
that binds the capsule's file names, a timestamp token that must verify before it counts as a
timestamp, and mocked replay that serves recorded MCP and declared-tool results
(Appendix~\ref{app:changes}). They have not been re-evaluated in the companion paper's
experimental setup; that re-evaluation, in particular of tool replay, is future work.

\paragraph{What ships at v\nfVersion.} The \cmd{nova} CLI exposes \nfTopCmds{}
top-level commands and \nfCmdPaths{} runnable command paths. Behind it sit \nfPackages{}
Python subpackages, about \nfPyKLOC{} thousand lines, exercised by \nfTestFuncs{} test
functions under a 90\,\% coverage gate. Further surfaces:
\begin{itemize}
  \item a local dashboard with \nfDashTabs{} tabs over \nfServeRoutes{} HTTP routes, each
        route classified in a deny-by-default authorization table;
  \item a multi-user server mode with OIDC, RBAC, SAML, SCIM and offline tokens;
  \item a Go collector for cluster ingestion, and a TypeScript SDK;
  \item \nfAdapters{} framework adapters, \nfRunners{} execution runners and \nfExportCmds{}
        evidence exporters for regulatory and archival formats.
\end{itemize}
\nfADRs{} Architecture Decision Records document the reasoning. The maintainers keep them in a
private design repository, so this count, unlike the others, cannot be re-derived from public
sources (Appendix~\ref{app:method}); the behaviour they describe is in the code, the public
\fpath{docs/} tree and \fpath{CHANGELOG.md}.

\paragraph{What is new in this edition.} The engines matured more than the feature list
grew. Since Edition~1:
\begin{itemize}
  \item \textbf{Mocked replay fails closed.} Any divergence ends the replay with a
        non-zero exit code (ADR-0300). SDK wire records are de-duplicated into logical
        calls (ADR-0305). Declared Python tools can now be substituted during replay
        (ADR-0306).
  \item \textbf{\cmd{nova diff} has a stable exit-code contract} and compares recorded
        request parameters (ADR-0303).
  \item \textbf{NovaSeal reports how far a signer's identity can be trusted}
        (\code{identity\_trust}). \cmd{nova seal init} creates a local CA (ADR-0301), and
        the verifier checks TSA chains and CRLs offline.
  \item \textbf{The dashboard server authorizes every route} (ADR-0228). It declares
        its tenancy posture (ADR-0229), ships with a secure-by-default Helm chart
        (ADR-0230), and writes audit records that do not overstate who acted (ADR-0231).
  \item \textbf{Trust claims are checked end to end} (v0.106). The signing intent is
        inside the signature, the capsule root commits to file names, \cmd{nova verify}
        checks an Evidence Bundle's signatures and subjects, maker-checker signers must
        chain to an operator-held trust list or CA, and batch jobs no longer inherit the
        submitting shell (ADR-0309--0315).
  \item \textbf{One verifier behind every trust surface} (v0.107). \cmd{nova verify}, the
        dashboard's trust radar and Seal tab, and the verify routes share one verifier for
        capsules and one for Evidence Bundles. Every bundle member is signed. Links, FIFOs
        and devices that a workload plants in its capsule are removed and recorded, never
        followed. Every route into production is gated: registration accepts only
        \code{development}, HTTP, \cmd{nova server} and prompt promotion pass the
        \code{promote\_direct} gate, and so does a rollback unless the target was in
        production before. The release also changes behaviour in many places;
        Appendix~\ref{app:changes} points to the list.
  \item \textbf{All four at-scale lineage backends are implemented} and
        container-verified: Postgres, KuzuDB, Apache AGE and JanusGraph.
  \item \textbf{A record-only evidence layer} covers human oversight, safety cases,
        energy receipts, scientific reproducibility and multi-agent hand-offs.
\end{itemize}
Appendix~\ref{app:changes} has the full delta.

\begin{keyidea}[How to read this document]
Every capability is labelled \works{} (implemented, tested, released), \exper{}
(released, but interface or scale validation may change), \planned{} (roadmapped) or
\fdesign{} (documented intent, no implementation). Statements about the code cite the
package or ADR that backs them. ``Honest limits'' boxes state where the system stops, so a
reader never has to infer it.
\end{keyidea}

\chapter{The problem and the idea}\label{ch:problem}

\section{Agents became behaviours}
A modern agent run assembles prompts from registries and calls foundation models whose
weights and routing change without notice. It invokes tools that mutate real systems,
reads and writes files, and spawns sub-agents across processes and machines. Four
questions follow every incident, and with conventional tooling they are mostly
unanswerable:

\begin{enumerate}
  \item \textbf{What exactly did the agent do?} Not what a framework chose to log, but
        what crossed the network, to which endpoint, under which environment.
  \item \textbf{Can the record be trusted?} A log anyone can edit is not evidence. Without
        cryptographic integrity, a tampered trace looks exactly like a genuine one.
  \item \textbf{Can it be inspected or reproduced later?} Providers drift. ``Run it
        again'' is a new experiment, not a reproduction.
  \item \textbf{What changed, and what did it affect?} Regression detection across model
        versions, and blast-radius analysis when a bad asset is found, need structured
        records and a lineage graph.
\end{enumerate}

\begin{nftable}
\begin{tabularx}{\linewidth}{@{}P{0.30\linewidth}L@{}}
\toprule
\thead{Who} & \thead{The moment it hurts} \\
\midrule
ML engineer, flaky production agent & ``It worked yesterday, it fails today, and I cannot replay yesterday.'' \\
Researcher moving laptop {\sffamily ↔} HPC & ``My Slurm run differs from my laptop run and I cannot see why.'' \\
Compliance officer & ``The auditor asked which model processed this person's data on 15 March.'' \\
Platform engineer & ``Which deployment, which agent, which prompt version caused this regression?'' \\
\bottomrule
\end{tabularx}
\end{nftable}

\section{The idea: one primitive}
NovaFabric's answer is deliberately narrow:

\begin{keyidea}
Every important AI-agent execution becomes a \textbf{sealed, replayable, diffable,
auditable evidence capsule}. If a system called a model, used a tool, read data or made
a decision yesterday, a developer can load that run today, or in two years. They can
re-execute it under a stated replay contract, compare it with another run, and export a
signed bundle for an auditor.
\end{keyidea}

Around that primitive the project invents exactly two top-level formats: the Run Capsule
and the Evidence Bundle. It aligns everything else with existing open standards, so the
record stays portable and customer-owned, and other tools can read it.

\section{Design commitments}
\begin{itemize}
  \item \textbf{Local-first now, distributed-ready always.} No core feature needs a
        server, a network connection or a cloud account. The cluster tier adds
        infrastructure \emph{on top of} the same capsule format without changing it.
  \item \textbf{Capture never blocks the workload.} By default every capture layer fails
        open: an error inside NovaFabric is logged and swallowed, never raised into the agent.
        The one exception is the opt-in strict capture mode, which refuses overlapping captures.
  \item \textbf{Framework-neutral, wire-level by default.} Capture intercepts SDK and
        HTTP calls, so it observes what was sent rather than what a framework reported.
        Adapters use each framework's own extension points.
  \item \textbf{Standards inside, few verbs outside.} OTel GenAI, OpenLineage, PROV,
        DSSE/in-toto, RFC~3161, RFC~8785, OPA/Rego and CycloneDX sit behind a small set of
        memorable commands.
  \item \textbf{Honesty as an engineering property.} Maturity labels are part of the
        documentation contract, guard tests pin documented counts to the code, and a
        withdrawn claim is withdrawn in writing (Chapter~\ref{ch:performance}).
\end{itemize}

\section{What NovaFabric is not}
The non-goals are written down and enforced in design review. NovaFabric is not an
orchestrator or scheduler, a training platform, an inference server or an LLM gateway.
It is not a hosted SaaS, a vector database, a labelling tool, or a real-time monitoring
and alerting product. It produces \emph{evidence for} compliance processes and never
claims that an export \emph{is} compliance.

\chapter{Landscape and differentiation}\label{ch:landscape}

\section{Four adjacent families}
Each neighbouring family of tools solves one slice of the problem well:

\begin{nftable}
\begin{tabularx}{\linewidth}{@{}P{0.20\linewidth}P{0.20\linewidth}P{0.22\linewidth}L@{}}
\toprule
\thead{Family} & \thead{Representatives} & \thead{Solves} & \thead{Leaves open} \\
\midrule
LLM observability & LangSmith~\cite{langsmith}, Langfuse~\cite{langfuse} & tracing, dashboards, evals & mutable records in a service database; no replay contract; no cryptographic integrity \\
Tracing standards & OpenTelemetry GenAI & vendor-neutral spans & no capsule or replay concept; no integrity; no data-to-model lineage \\
Supply-chain attestation & in-toto~\cite{intoto}, SLSA~\cite{slsa}, Sigstore~\cite{sigstore} & signed build provenance & no agent-runtime semantics, capture or replay \\
Data lineage & OpenLineage, Marquez & pipeline lineage & no model-call or agent-run semantics \\
\bottomrule
\end{tabularx}
\end{nftable}

Tracing answers \emph{what happened}. NovaFabric asks a stricter question: \emph{can
this run be safely re-examined, and can the record be trusted?} It combines the slices
into one portable record and consumes or emits each family's standard, so it slots into
an existing stack instead of replacing it.

\section{An executable differentiation table}\label{sec:difftable}
The project keeps a ten-claim differentiation table, and \fpath{scripts/verify\_differentiation\_table.py}
checks every claim against the code. A test (\fpath{tests/test\_differentiation\_table.py})
runs the script and fails if any claim stops holding. At v\nfVersion{} it reports
\textbf{\nfDiffPass{} of \nfDiffClaims{} claims PASS}.

\begin{nftable}
\begin{tabularx}{\linewidth}{@{}lL@{}}
\toprule
\thead{ID} & \thead{Claim (as checked)} \\
\midrule
D-01 & Self-contained: registry and capsule writer work with no network or cloud service \\
D-02 & Cryptographic evidence: DSSE envelopes, Merkle log, pluggable signing backend \\
D-03 & Replay contract with named modes: forensic, mocked, semantic, exact \\
D-04 & Structured cross-run diff as a first-class API (\code{DiffReport}) \\
D-05 & Policy gates plus maker-checker dual approval on promotion (OPA/Rego) \\
D-06 & WORM evidence storage adapters (S3 Object Lock, Azure immutable blobs, GCS Bucket Lock) \\
D-07 & Versioned capsule schema with a migration CLI \\
D-08 & OpenLineage emission from the lineage graph \\
D-09 & Multi-framework adapters through native extension hooks \\
D-10 & Eval-gated promotion (Rego regression gate) \\
\bottomrule
\end{tabularx}
\end{nftable}

\begin{honest}[What the check proves]
The verifier is a \emph{structural} check: the mechanism behind each claim exists and
imports. It is not a behavioural benchmark against competitors. The behaviour is covered
by the test suite (Chapters~\ref{ch:replay} to~\ref{ch:governance}), and the comparison
with other tools is the author's reading of their public documentation.
\end{honest}

\section{Related research}\label{sec:related}
Three research lines sit close to the capsule. \emph{Provenance capture} systems record what a
computation did below the application: PASS~\cite{pass2006}, SPADE~\cite{spade2012},
Hi-Fi~\cite{hifi2012}, Linux Provenance Modules~\cite{lpm2015}, CamFlow~\cite{camflow2017} and
ProTracer~\cite{protracer2016} capture whole-system provenance, and Davidson and
Freire~\cite{davidsonfreire} survey the workflow side. NovaFabric records at the model, tool
and file boundary of an agent run; it does not replace system-level provenance and does not
compete with it. \emph{Tamper-evident logging} supplies the primitives that NovaSeal
composes: hash-linked timestamping~\cite{haberstornetta}, audit logs with forward
integrity~\cite{schneierkelsey} and Merkle-tree logs~\cite{crosbywallach,rfc6962}. The
contribution is their integration at the granularity of an agent session, not new
cryptography. \emph{Agent observability and accountability} is a recent and mostly
proposal-stage literature: AgentOps~\cite{agentops}, AgentSight~\cite{agentsight} and
AgentTrace~\cite{agenttrace} address observing LLM agents, a trace-based assurance
framework~\cite{traceassurance} and Auditable Agents~\cite{auditableagents} address assurance
and auditability, and blockchain-anchored evidence~\cite{agentblackbox} and a replay
harness for financial agents~\cite{replayablefinancial} address evidence and replay. These
works are cited as context. This report does not benchmark NovaFabric against them, and a
comparison on shared workloads is future work.

\section{Positioning}
NovaFabric complements agent runtimes, clouds and observability platforms. It competes
only at the provider-neutral evidence, replay and audit layer. The strongest external
need it serves is a \emph{neutral, portable, customer-owned, tamper-evident} record of
agent behaviour across models, tools, clouds and time. A provider that records its own
ecosystem is not an independent witness of it; that gap is what this layer targets.

\chapter{Concepts and data model}\label{ch:concepts}

\section{Five primitives}
Everything NovaFabric records reduces to five composable primitives:

\begin{nftable}
\begin{tabularx}{\linewidth}{@{}lP{0.32\linewidth}L@{}}
\toprule
\thead{Primitive} & \thead{Definition} & \thead{Examples} \\
\midrule
\textbf{Node} & something that exists & a run, an agent, a model, a tool, an asset \\
\textbf{Edge} & a typed relationship & run \emph{consumed} dataset; run B \emph{replayed\_from} run A \\
\textbf{Event} & something that happened & a model call, a tool call, a policy decision, a file write \\
\textbf{Artefact} & immutable, content-addressed bytes & a request payload, a media output, an SBOM \\
\textbf{Seal} & proof the bytes were not changed & DSSE signature + RFC~3161 token + Merkle inclusion \\
\bottomrule
\end{tabularx}
\end{nftable}

Identifiers follow one strategy. Runs, events and lineage edges use time-sortable ULIDs,
generated in \fpath{capture/\_ulid.py}; immutable artefacts are identified by the SHA-256 of
their bytes. Re-importing a capsule replaces that run's lineage edges, so the graph comes out
the same. Provider identifiers are stored but never used as primary keys.

\section{The Run Capsule}\label{sec:capsule}
A captured run is a \textbf{directory of flat files}, never a database row
(Figure~\ref{fig:fig-05-capsule-anatomy}). The capsule is the source of truth. The
registry, the metadata store, the lineage graph and the knowledge graph are all derived
indexes that \cmd{nova} can rebuild by re-reading capsules. Losing an index is a recovery
task; losing capsules is a disaster.

\nffigure{fig-05-capsule-anatomy}{Anatomy of a Run Capsule at v\nfVersion. Dashed boxes are written only when observed or enabled.}

The capsule is written in a fixed order by the shared finalize path
(Chapter~\ref{ch:capture}). Once the workload has exited, the capsule is treated as
untrusted input: a sweep removes every symbolic link, FIFO, socket and device found at any
depth, without following it, and records each removal as tampering. Content is then
redacted \emph{before} \code{capsule.yaml} is
written. A residual rescan follows, and its result goes into
\code{redaction-proof.json}. Finally \code{capsule.yaml} is rewritten with
\code{evidence\_digests}, a digest of every evidence file. When the capsule is sealed,
\cmd{nova verify} checks that \code{capsule.yaml} matches the signed payload and that
every file matches its digest, so an edit to any evidence file after sealing is detected.
A manifest swapped for an edited \code{capsule.json}, and a link or special file anywhere in
the capsule, also fail verification.
On an unsealed capsule the digests are unauthenticated and nothing checks them yet.
When events were dropped under back-pressure, or a recorder failed and was swallowed,
\code{capture-health.json} says so. It is
itself covered by the digests, so the capsule records its own gaps.

\begin{honest}[Schema versions and content]
\begin{itemize}
  \item The manifest in force is \code{schema\_version:}~\code{"0.1.0"}
        (\fpath{src/novafabric/schemas/run-capsule.schema.json}). The v1.0 capsule schema
        in \fpath{schemas/} is the \emph{target} of the format freeze and is marked
        ``not yet in force'' (Chapter~\ref{ch:roadmap}).
  \item v0.105 adds fields (\code{host.runner}, a residual-check block). As a result,
        \cmd{nova validate} from v0.104 or earlier rejects capsules written by v0.105.
        Readers must be at least as new as the writer. v0.107 adds three optional stream
        references to the schema in the same way (\code{network\_events\_ref},
        \code{file\_events\_ref}, \code{human\_approvals\_ref}).
  \item As built, model-call request and response content is stored at every capture
        level, and \code{minimal} behaves like \code{standard}: only \code{forensic} and
        \code{air\_gapped} change what is recorded. Treat a capsule as holding full prompts
        and responses. Tool payloads
        recorded through \code{record.tool} and the ADK and OpenAI Agents adapters are kept
        in full only at the \code{forensic} and \code{air\_gapped} levels. The MCP hook
        always records full arguments and results, because mocked replay serves them.
\end{itemize}
\end{honest}

\section{The wire contract: Event Envelope v1}
All evidence events conform to one normative, frozen wire format, whether they come
from a laptop SDK or from a collector fleet. Its status is \emph{accepted}, and a
breaking change requires \code{envelope\_version: "2"}, never an in-place edit. The
envelope carries:
\begin{itemize}
  \item \textbf{identity:} \code{event\_id}, \code{global\_run\_id}, \code{run\_id},
        \code{parent\_run\_id};
  \item \textbf{location:} cluster, tenant and emitter node;
  \item \textbf{causality:} W3C trace context;
  \item \textbf{integrity:} a \code{payload\_hash} that commits to the payload without
        carrying it inline.
\end{itemize}
It ships as JSON Schema, a proto3 definition with deterministic-encoding rules for
signing, normative OTLP and CloudEvents mappings, and a SHA-256 pin that the test suite
verifies (\fpath{schemas/event-envelope-v1/}). Since the freeze the \code{event\_type}
enum has grown twice, both times additively and with the pin updated: the canonical
\code{Run*}, \code{ModelCall*} and \code{ToolCall*} lifecycle events (ADR-0220).

\section{Distributed identity: parent and child capsules}
A distributed job becomes a tree of capsules under one \code{global\_run\_id}. A
Kubernetes job, a Slurm array or an MPI training run all fit this shape. The as-built
schema (\fpath{capsule/schema.py}) defines:
\begin{itemize}
  \item the capsule role (\code{PARENT}, \code{CHILD}, \code{STANDALONE}) and a separate
        distribution role (\code{DRIVER}, \code{WORKER});
  \item parent states \code{RUNNING}, \code{COMPLETE}, \code{PARTIALLY\_COMPLETE} and
        \code{FAILED};
  \item a fail mode, \code{fail-open} by default, that decides whether a parent missing
        some of its expected children finalizes as \code{PARTIALLY\_COMPLETE} or
        \code{FAILED}.
\end{itemize}
The arrival policy is designed for distributed reality, where children may arrive
before their parent (Chapter~\ref{ch:scale}).

\section{Typed lineage vocabulary}
Capture writes \code{consumed}, \code{produced\_by} and \code{replayed\_from} edges into
\code{lineage.jsonl}. The distributed vocabulary in \fpath{capsule/schema.py} adds
\code{contains}, \code{spawned}, \code{delegated\_to}, \code{member\_of\_session},
\code{wrote\_memory} and \code{read\_memory}, each mapped to W3C PROV-DM where a
counterpart exists (ADR-0044). Sessions use a deliberately separate \emph{grouping} edge,
so a sequence of independent runs is never mistaken for one distributed job. \cmd{nova migrate}
additionally writes a \code{migration\_source} edge (new in v0.107) that names the original
capsule by run id and manifest digest.

\section{Maturity vocabulary}
\begin{nftable}
\begin{tabularx}{\linewidth}{@{}lL@{}}
\toprule
\thead{Label} & \thead{Meaning} \\
\midrule
\works & implemented on \texttt{main}, tested, shipped in a released version \\
\exper & shipped, but the interface or format may change, or scale validation is pending \\
\planned & on the roadmap with a target; not yet implemented \\
\fdesign & documented design intent; no implementation and no committed target \\
\bottomrule
\end{tabularx}
\end{nftable}

\chapter{Architecture}\label{ch:architecture}

\section{System context}
NovaFabric sits between three kinds of party. On one side are people: developers,
operators and auditors. In the middle is the workload it wraps. On the other side are
the external systems that workload touches, and the systems that consume NovaFabric's
evidence. It observes and records; it never plans or schedules the workload.

\nffigure{fig-02-system-context}{System context. NovaFabric wraps the workload and observes
its calls at the wire. It talks only to the services an operator configures: timestamp
authorities, signing services, object storage and lineage or SIEM sinks.}

\section{Subsystem map}
The source tree has \nfPackages{} Python subpackages under \fpath{src/novafabric/}, about
\nfPyKLOC{} thousand lines in total. Two further components sit outside it: a Go
collector (\fpath{collector/}) and a TypeScript SDK (\fpath{packages/nova-sdk-ts}). The
packages group into eight domains (Figure~\ref{fig:fig-03-subsystem-map}). Each name in
the figure is a real package.

\nffigure{fig-03-subsystem-map}{Subsystem map: eight domains over \nfPackages{} packages
(representative packages shown). Capture writes capsules. Everything else asks questions
of the record or serves it.}

\begin{nftable}
\begin{tabularx}{\linewidth}{@{}P{0.24\linewidth}L@{}}
\toprule
\thead{Domain} & \thead{Responsibility, and what is notable as built} \\
\midrule
Capture & wire, SDK and framework interception; runners; masking; warm daemon. Largest single package: \code{capture} \\
Evidence model \& storage & capsule hierarchy, sessions, envelope, asset specs and registry, WORM and object storage, derived metadata index \\
Analyse & replay (five modes), structural diff, failure attribution, drift detectors, lineage and the capsule knowledge graph, offline query DSL \\
Trust \& integrity & NovaSeal (DSSE, RFC~3161, Merkle log), evidence bundles, maker-checker, preservation, evidence-grade backup \\
Govern \& comply & regulatory exporters, PII erasure key lifecycle, retention and legal holds, hash-chained audit log, cost attribution \\
Assurance \& safety & eval suites and score cards, OWASP LLM checks, safety-case compiler, human-oversight and frontier-safety evidence \\
Domain facets & record-only evidence facets attached to a capsule: A2A hand-offs, context and retrieval provenance, memory, embodied, science, settlement \\
Serve \& scale & CLI, local dashboard API, multi-user server, Python client, durable jobs, writer lease, evidence fabric, collector app \\
\bottomrule
\end{tabularx}
\end{nftable}

\section{Design invariants}
Eleven invariants are written into the developer architecture guide
(\fpath{docs/architecture.md}). Reviews treat them as load-bearing. Each one backs a
promise made to users:

\begin{enumerate}
  \item A capsule is portable and self-describing. It verifies with no server, network
        or database.
  \item Capture never blocks the workload by default. If a NovaFabric component fails,
        the workload continues and the failure is recorded.
  \item Secrets are meant to stay inside the redaction boundary. This is a design
        objective, not a guarantee: the scanner matches known patterns only, and its
        documented blind spots are listed in Chapter~\ref{ch:capture}.
  \item Full prompt and response capture should be opt-in. \textbf{As built, this
        invariant is not met}, and the guide says so (see the box below).
  \item No silent telemetry and no update checks, ever.
  \item Core local-mode features require no internet.
  \item Compute-node hot paths never write to a database or graph. They write to a local
        spool, and something else ingests it.
  \item Schema changes are additive and optional first.
  \item No root containers and no privileged Kubernetes access.
  \item Exactly two top-level formats: Run Capsule and Evidence Bundle.
  \item CLI and route-table structure is introspected by shape, never by class identity
        (ADR-0250). Twice, a framework upgrade made an \code{isinstance} walk report an
        empty surface for a system that worked. The fact probes behind this document use
        the same shape-based walkers.
\end{enumerate}

\begin{honest}[Invariant 4]
A default \cmd{nova capture} of a Python workload writes request messages and response
choices into \code{model-calls.jsonl}, and mocked replay depends on those stored
responses. \code{--capture-content} only governs what the OpenTelemetry span export
carries. Changing the default is an open design decision. Until it changes, deploy
NovaFabric as a system that stores prompts. Its redaction, masking and WORM controls exist
for exactly that reason. The capture-level policy does not yet reduce what model calls
store: \code{minimal} and \code{standard} behave alike, and its per-field allowlist is
applied by no capture path (\planned).
\end{honest}

\section{Two tiers, one format}
The \emph{local tier} runs a single process with SQLite indexes and capsules on disk. It
covers capture, seal, replay, diff, lineage, policy and export without any service. The
\emph{cluster tier} adds a Go node collector, a message bus, an object capsule store, a
Postgres metadata index under row-level security, and graph backends. It adds them
around the same capsule and envelope formats (Chapter~\ref{ch:scale}). Because the
cluster tier changes where evidence is stored and indexed, not what it is, a capsule
captured on a laptop can be verified, replayed and diffed against one captured on a
cluster.

\chapter{Capture runtime}\label{ch:capture}

\section{Capture at the wire}
NovaFabric captures what actually crossed the boundary, not a framework's self-report.
SDK hooks for OpenAI, Anthropic and MCP record logical calls. Wire hooks for
\code{httpx}, \code{requests}, \code{aiohttp} and \code{urllib3} record transport
attempts; an optional \code{httpx2} hook, installed only when that package is importable,
covers the client that \code{openai} 3.x builds. An OpenTelemetry GenAI hook and a Python-tool hook complete the set
(\fpath{capture/hooks/}). Every entry point converges on one recorder and one finalize
path (Figure~\ref{fig:fig-04-capture-flow}).

\nffigure{fig-04-capture-flow}{Capture data flow. Every entry point feeds
the same recorder and the same finalize path, so every capsule carries the same proofs
whichever way it was captured.}

\begin{nftable}
\begin{tabularx}{\linewidth}{@{}P{0.28\linewidth}LP{0.13\linewidth}@{}}
\toprule
\thead{Entry point} & \thead{Mechanism and when to use it} & \thead{Status} \\
\midrule
\cmd{nova capture <cmd>} & launches the command with hooks injected; zero code change. \code{--runner} picks where it executes & \works \\
SDK \code{@agent} / \code{record} & in-process decorator and façade; \code{@record.tool(mutation\_class=\dots)} declares tools for replay & \works \\
\nfAdapters{} framework adapters & LangGraph, AutoGen, CrewAI, DSPy, LlamaIndex, Pydantic AI, Haystack, OpenAI Agents, Google ADK, Bedrock AgentCore, A2A, each through the framework's own hook & \exper \\
\cmd{nova mcp-proxy} & transparent stdio and HTTP+SSE MCP proxy for agents that cannot be instrumented & \works \\
\cmd{nova api-proxy} & base-URL override proxy at the egress. TLS interception is permanently rejected & \works \\
\cmd{nova daemon} / \code{novacap} & warm pre-fork \code{AF\_UNIX} capture daemon, Linux-only and opt-in; a busy or refusing daemon makes \code{novacap} run \cmd{nova capture --no-daemon} instead, and a run the daemon accepted is never repeated & \exper \\
OTLP ingest, \cmd{nova ingest-capsule}, \cmd{nova import} & build or accept capsules from OTel spans, existing capsules or verified batches & \exper \\
\bottomrule
\end{tabularx}
\end{nftable}

\paragraph{Logical calls, not wire attempts (ADR-0305).} One SDK call can produce several
HTTP attempts: retries, redirects, streaming. Since v0.105 those wire records are tagged
\code{record\_role: transport} with a shared \code{logical\_call\_id}. Every consumer
counts logical calls only: cost, diff, replay, query, the knowledge graph and the
exporters. A capsule written before v0.105 is read through a fallback. Its
\emph{sealed} \code{model\_call\_count} keeps the doubled figure, so the count should
be re-derived from \code{model-calls.jsonl}.

\section{Fail-open, bounded, isolated}
Every capture layer fails open. An exception inside NovaFabric is logged and recorded,
never raised into the workload. Since v0.107 that holds at every record site of the
OpenAI, Anthropic, MCP and wire hooks: a recorder failure (an SDK object that does not
serialise, a full disk) is logged as \code{capture.record\_dropped} (once per hook,
stream and error type) and counted in \code{capture-health.json}, and the call's own
return value or exception passes through unchanged. A hook never consumes a streamed
response: the \code{requests} hook once read the body to size the network event, so an
SSE call returned only when the stream ended. \code{response\_size\_bytes} is now the
size of an already-read body, else \code{Content-Length}, else \code{null}. A streamed
SDK call records only into the capture that opened it; a record that arrives after that
capture ended is dropped and counted (\code{capture.stream\_record\_dropped}), never
written into another capsule. Queues are bounded. When events are dropped under
pressure, \code{capture-health.json} records it, and the digests cover that file too. Since
v0.102, concurrent captures in one environment are isolated from each other. Since
v0.103, a strict capture mode lets an operator choose to refuse rather than degrade.

\section{The workload is untrusted input}
Under \cmd{nova capture} the workload writes into its own capsule directory, and with
\code{--runner docker} (or Kubernetes, through \code{kubectl cp}) that directory is
mounted read-write into a container. Since v0.107 finalisation therefore treats the
capsule as untrusted once the workload has exited. A sweep removes every symbolic link,
FIFO, socket and device at any depth without following it. Each removal is recorded in
\code{capture-health.json} (\code{tampered\_entries}), in \code{redaction-proof.json}
and in the manifest (\code{metadata.capture\_tampered}); \cmd{nova verify} and
\cmd{nova validate} then fail the capsule, while capture itself still exits with the
workload's own code. Every later read of a capsule file refuses links and never blocks on
a FIFO. Every later rewrite, by the secret scanner, the masking pipeline or the lineage
and capture-health writers, replaces the file by rename instead of writing through it.
Before this, a planted link could make the scanner rewrite a file outside the capsule,
and a FIFO could hang capture. Hard links are not treated as tampering, and remain a
documented limit.

Schema validation of tool arguments and results follows the same rule. It never
retrieves a \code{\$ref}: a nested remote reference, an \code{\$id}-relative one, a
\code{\$dynamicRef} or a \code{file://} reference used to make capture send an HTTP
request or read a local file at every capture level, \code{air\_gapped} included. Only
references inside the schema document and the bundled meta-schemas resolve. Anything else
gives a \code{null} verdict with \code{schema-unresolved}.

\section{Capture levels}\label{sec:capture-levels}
Four levels exist in \fpath{policies/capture\_level.py}: \code{minimal} (metadata only),
\code{standard} (the default), \code{forensic} (full payloads) and \code{air\_gapped}
(full payloads; meant to forbid external calls, which it does not do yet, see the box below).
The level is read from \code{NOVA\_CAPTURE\_LEVEL} and nowhere else. \cmd{nova policy capture-level set}
validates a level and prints the variable to export; it persists nothing.

\begin{honest}[Where capture levels apply today]
\exper{} The level separates exactly two groups: \code{forensic} and \code{air\_gapped}
against the rest. At those two, the \code{record} façade, the Python-tool hook and some
framework adapters (LangGraph, Google ADK, OpenAI Agents) also keep payloads: tool
arguments and results, state, document text. Otherwise they keep digests.
\code{minimal} and \code{standard} behave identically. The per-field allowlist that
describes \code{minimal} as metadata-only is applied by no capture path, and
\code{air\_gapped} blocks no network call (both \planned). The level does not govern the
SDK and wire hooks, which is why model-call content is stored at every level
(Chapter~\ref{ch:architecture}, invariant~4). The same holds for MCP tool calls. Tool
payloads from \code{record.tool} and the ADK and OpenAI Agents adapters are gated: below
\code{forensic} only post-redaction digests are kept.
\end{honest}

\section{Secrets, masking and redaction proofs}
Before any manifest is written, the finalize path scans and redacts every capsule file
that ADR-0009 names (complete since v0.104) with the \code{gitleaks-core} rule pack. A
\emph{residual rescan} then checks the already-redacted output (v0.105) and records the
result in \code{redaction-proof.json}. A pluggable, fail-closed masking pipeline
(\code{masking}, ADR-0135) adds PII masking through \code{--masker}.

\begin{honest}[Pattern-based scanning]
The secret scanner is rule-based. Its documented blind spots include PEM keys, JWT and
bearer tokens, passwords, connection strings, several cloud-vendor key formats and
personal data in general. \fpath{docs/architecture/run-capsule.md} lists them, and a
test keeps that list in step with the rule pack. Use the masker for data that the pattern
pack does not cover. Capture levels do not help here: model-call content is stored at every
level (Section~\ref{sec:capture-levels}), so a prompt that holds sensitive data is protected
by neither the scanner nor the level.
\end{honest}

\section{Runners: where a captured command executes}
One protocol (ADR-0025) abstracts \nfRunners{} execution targets: \code{local} (the
default), \code{docker}, \code{kubernetes} (a Job through \code{kubectl}), \code{slurm}
(\code{sbatch}/\code{sacct}), \code{pbs} and \code{lsf}. Since v0.105 the manifest records
\code{host.runner}, the OCI image digest and the Slurm scheduler context (ADR-0307). The
Slurm runner can also attach \code{sacct} energy receipts. No off-host runner forwards the
submitting shell. The Docker and Kubernetes runners pass an allowlisted set of environment
variables (ADR-0270), which closed a leak in which every submit-side variable was written
into the Kubernetes Job object. Since v0.106 SLURM and LSF submit from a minimal
environment (ADR-0311), because \code{sbatch} and \code{bsub} copy their own environment
into the job by default, and no runner forwards a NovaFabric credential. Anything else a
job needs, such as an API key, is passed explicitly with the \code{extra\_env} runner
option. Since v0.107 such a value is no longer put on the command line of the submit-side
tool that carries it. SLURM
reads its batch script from \code{sbatch}'s standard input (it was \code{--wrap}). LSF pipes
its script to \code{bsub} from memory, where it once wrote \code{nova-<run>.lsf} into the
shared capsule directory. Docker passes \code{-e NAME} with the value in the client's own
environment (a name the Docker client itself reads keeps \code{-e NAME=value}). \code{slurm}, \code{lsf} and \code{pbs} quote every argument with
\code{shlex.quote}, so \code{\$}, backticks and \code{;} in a value reach the job
verbatim instead of being expanded by its shell. \code{extra\_env} also accepts a JSON
object, and a malformed one fails the run at setup, naming the key and never the value.
This is not secrecy end to end. The value is still in \cmd{nova capture}'s own argument
list, SLURM stores the batch script with the job, \code{docker inspect} shows it to anyone
who can reach the daemon, and the Kubernetes Job object carries it as a literal.
\chapter{Replay and diff}\label{ch:replay}

\section{Replay is a contract, not a button}
Hosted models drift. Re-running a prompt next month is a new experiment, so NovaFabric
does not claim byte-exact replay of remote LLM calls. Instead each replay mode states
exactly what it guarantees (\fpath{replay/\_flags.py}, \fpath{replay/\_replayability.py}):

\begin{nftable}
\begin{tabularx}{\linewidth}{@{}lP{0.17\linewidth}Ll@{}}
\toprule
\thead{Mode} & \thead{Re-executes?} & \thead{Guarantee} & \thead{Status} \\
\midrule
\code{forensic} & no & read-only inspection of the record; works on every capsule & \works \\
\code{semantic} & no & scores the recorded responses by text similarity (\code{difflib}), not by meaning & \works \\
\code{exact} & no & an eligibility report: a deterministic \code{env.lock} and a seed on every call & \works \\
\code{mocked} & yes & re-runs the recorded command with model and MCP results served from the capsule & \works \\
\code{intervention} & yes & counterfactual replay (ADR-0086); model-call substitutions reach the re-run workload, tool-call substitutions change only the output capsule & \exper \\
\bottomrule
\end{tabularx}
\end{nftable}

\nffigure[0.55\linewidth]{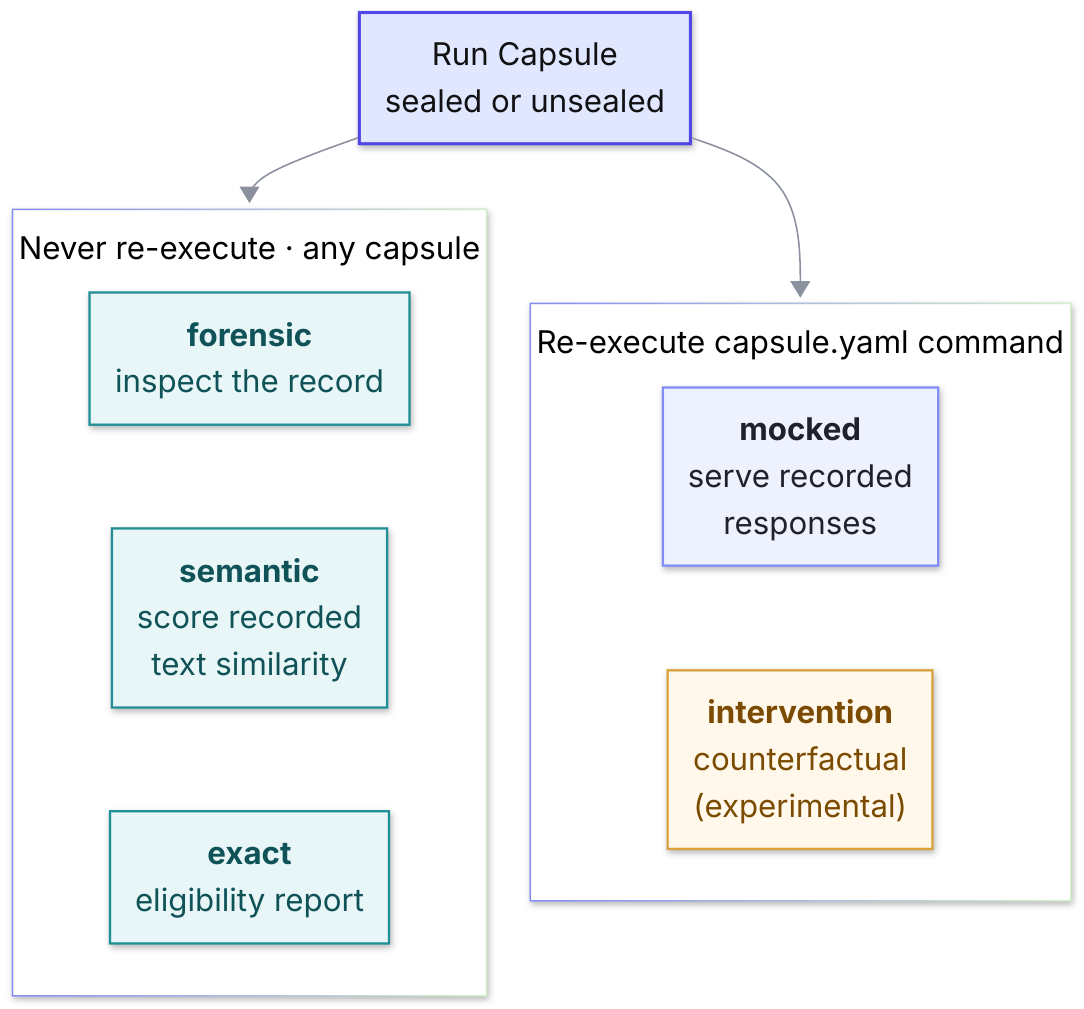}{Replay modes split by whether they re-execute
the recorded command. The three read-only modes work on every capsule.}

\section{The mocked-replay contract (ADR-0300, ADR-0304)}
Mocked replay installs an in-process dispatcher in the re-executed workload
(Figure~\ref{fig:fig-06b-mocked-contract}):
\begin{itemize}
  \item \textbf{Served from the capsule:} OpenAI Chat Completions and Responses, and
        Anthropic Messages, sync or async, streamed or not. Recorded SDK errors are
        re-raised. MCP \code{ClientSession.call\_tool} results are served too.
  \item \textbf{Refused, not approximated:} unsupported SDK surfaces, for example
        \code{messages.stream()}, \code{beta.messages} and raw-response wrappers. A
        support matrix is generated from \fpath{replay/\_support\_matrix.py}.
  \item \textbf{Fail-closed by default:} any divergence ends the replay with exit code 1
        and a \code{Replay\allowbreak{}Divergence} record. This covers a call with no recorded
        response, an unmatched tool call, and a recorded response left unconsumed.
        \code{--permissive} restores warn-and-continue and still records every divergence.
        Under \code{--permissive}, an unsupported model surface goes live to the real
        provider.
\end{itemize}
\nffigure[0.82\linewidth]{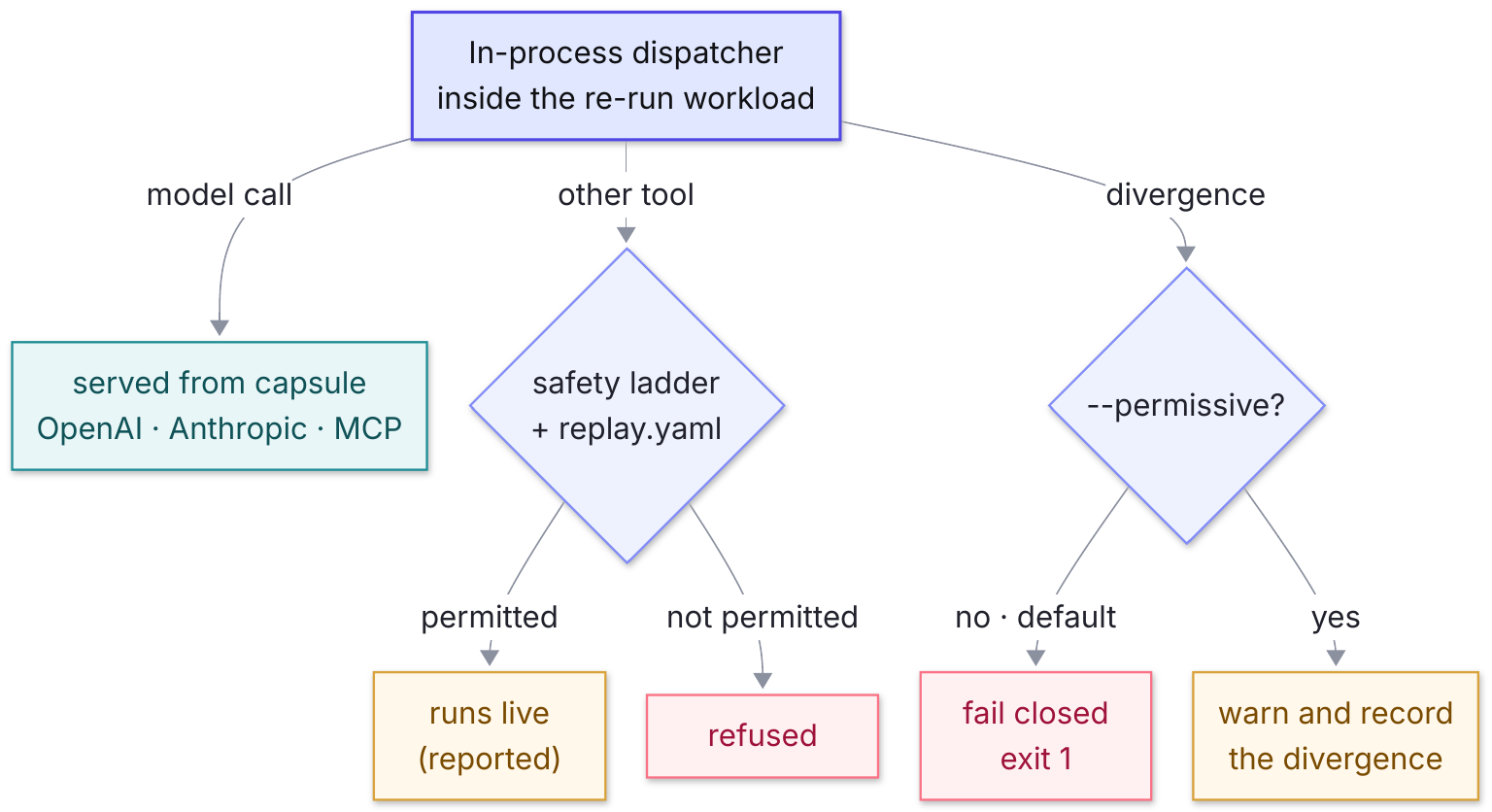}{The mocked-replay contract: each intercepted
call is served, run live or refused, and any divergence fails closed unless \code{--permissive}.}

Exit codes are part of the contract: 0 success; 1 failure or abort (unless the
replayed command itself exited non-zero, in which case its own code is returned); 2
environment mismatch, or a file the replay needs (\code{capsule.yaml}, \code{replay.yaml},
\code{env.lock}, a call stream) that is present but unusable, such as a link, a FIFO or a
directory, so nothing ran; 3 a tool override that cannot be enforced, so the replay refuses
to start; 124 the replayed command did not finish within 600\,s and was killed
(\code{ReplayTimeout}, which the CLI prints because a command can exit 124 itself);
otherwise the replayed command's own exit code.

\section{The safety ladder (ADR-0012)}
Anything not served from the capsule is classified by mutation class. It runs live only
if the operator passed the matching flag, and never against a \code{replay.yaml}
\code{allow: false}, even under \code{--permissive}. Since v0.106 the OPA
\code{replay\_mutating} policy is evaluated whenever any write-permitting flag is passed
(\code{--allow-mutating}, \code{--allow-external-side-effects} or
\code{--allow-unknown-mutation}). A deny, or a policy-engine error, refuses the replay
before anything runs:

\begin{nftable}
\begin{tabularx}{\linewidth}{@{}lL@{}}
\toprule
\thead{Mutation class} & \thead{Flag that permits it} \\
\midrule
\code{none} & always permitted \\
\code{read-only} & \code{--allow-readonly} \\
\code{idempotent-write}, \code{non-idempotent-write} & \code{--allow-mutating} \\
\code{external-side-effect} & \code{--allow-external-side-effects} \\
\code{unknown} (every MCP call is gated as unknown) & \code{--allow-unknown-mutation} \\
\bottomrule
\end{tabularx}
\end{nftable}

\section{Substituting non-MCP tools (ADR-0306)}
\exper{} Python tools declared with \code{@record.tool(mutation\_class=\dots)} can be
substituted from the capsule, and \code{replay.yaml} \code{tool\_overrides} are enforced
inside the replayed process (slices 1--2, v0.105). Since v0.106 Google ADK
(\code{make\_tool\_plugin}, a documented hook) and OpenAI Agents (\code{wrap\_tools}) tool
calls are captured and served too (slices 3--4), along with nested \code{record.tool}
boundaries and MCP matching after redaction.

\begin{honest}[Replay boundaries]
\begin{itemize}
  \item Capsules captured \emph{inside} a framework call (\code{capture\_mode:
        sdk-decorator}) or imported from OTel spans record no runnable command. Mocked
        replay refuses them up front with a precise reason (ADR-0306, open question 7).
        Intervention replay does not refuse them: it emits the counterfactual streams
        without re-executing and says so in the result. Forensic and semantic replay
        work as usual.
  \item \code{record.tool} and adapter tool results captured below the \code{forensic} level hold digests, not
        payloads, so they cannot be served (\code{tool\_result\_not\_servable}).
  \item The OpenAI Agents seam (\code{on\_invoke\_tool}) is undocumented upstream and
        has been verified against \code{openai-agents} 0.17.x only.
  \item Network connections during replay are reported, not blocked.
\end{itemize}
\end{honest}

\section{Structural diff and the CI gate (ADR-0303)}
\cmd{nova diff} compares two capsules structurally, not textually
(\fpath{diff/\_align.py}, Figure~\ref{fig:fig-07-diff-gate}). Model calls are paired in
three steps:
\begin{enumerate}
  \item by \code{parent\_span\_id}, when it is unique on both sides;
  \item by longest-common-subsequence anchors on identical requests;
  \item by position between those anchors.
\end{enumerate}
A changed prompt therefore shows as one changed pair, not as one call removed and
another added. Tool calls pair by name and argument hash, then by position among calls
with the same name. Since v0.105 every recorded request parameter (temperature, seed,
\code{top\_p}, \dots) is compared, and a change marks the pair \code{request\_changed}.

\nffigure{fig-07-diff-gate}{Diff alignment and the exit-code contract that makes \cmd{nova diff} a CI gate.}

The exit code is a published contract. Exit 1 is reserved for ``the comparison ran and
found a difference'' under \code{--assert-no-regressions} or \code{--assert-same-shape}.
Exit 2 means the capsules could not be compared at all, for example an unresolved
reference, an unreadable or malformed \code{capsule.yaml} or \code{env.lock}, or a record
file that is a symbolic link or a FIFO, which the diff never follows (v0.107). Exit 3 means the
\code{--significance} gate (a sequential probability ratio test) accepted the
regression hypothesis. A pipeline can therefore tell ``the model changed'' apart from
``the baseline is missing''.

\section{Drift, equivalence and the assurance loop}
\exper{} Offline detectors (\cmd{nova drift collect|\allowbreak{}detect|\allowbreak{}silent-failure|\allowbreak{}root-cause|\allowbreak{}fingerprint},
ADR-0147) gather evidence about drift and never rule on it: they exit 0 unless the input
is bad. The verdicts come from separate commands that exit 1 on failure:
\cmd{nova replay-equivalence check} (ADR-0144), \cmd{nova assure-baseline},
\cmd{nova assure-canary}, \cmd{nova assure-alarm} and \cmd{nova assure-run}. The split
is deliberate. An evidence system should separate collecting facts from deciding
policy, and NovaFabric never schedules the loop itself.

\chapter{Trust and evidence}\label{ch:trust}

\section{NovaSeal: three independent proofs}
NovaSeal (ADR-0041, \fpath{trust/novaseal/}) binds three proofs to a capsule's canonical
manifest. The manifest already carries \code{evidence\_digests} over every evidence
file, so tampering with any file breaks at least one proof
(Figure~\ref{fig:fig-08-trust-chain}). Since v0.107 the seal also records which manifest
file it was made over, so renaming \code{capsule.yaml} to an edited \code{capsule.json}
fails verification, and readers of a sealed capsule use the signed payload. One verifier
(\code{verify\_capsule}) now sits behind \cmd{nova verify}, the trust radar, the dashboard
Seal tab and \code{POST /api/runs/\{id\}/verify}, so a tampered or transplanted capsule can
no longer read \code{attested} on one surface while the command line fails it.

\nffigure{fig-08-trust-chain}{The trust chain. Who signed, when it was signed and where the
signature sits in history are proven independently and checked together by \cmd{nova verify}.}

\begin{enumerate}
  \item \textbf{Who: a DSSE signature.} Backends: local ECDSA P-256 (\works); AWS KMS,
        GCP KMS and Azure Key Vault (\exper); a mock backend for tests. Sigstore keyless
        signing (ADR-0071, \exper) is a separate path, deliberately outside the
        \code{SigningBackend} protocol. Cloud key-wrapping for envelope encryption is
        also available (ADR-0185, \exper).
  \item \textbf{When: an RFC~3161 timestamp token.} Multi-TSA fallback, a nonce store
        against replay, and offline verification of the TSA chain with CRL checks
        (ADR-0070, \exper). A token that cannot be verified is not a timestamp. Until
        v0.107 the fallback for a token that the strict parser rejects accepted an
        unverifiable signature, and one non-DER byte let a genuine token with a swapped
        imprint verify. It now requires a verifying TSA signature, a signed
        \code{messageDigest} equal to the hash of the \code{TSTInfo}, and the imprint read
        by position.
  \item \textbf{Where in history: a Merkle-log leaf.} An append-only log per tenant,
        in SQLite by default or Postgres. Append-only behaviour is enforced by database
        triggers and rules. Tree heads record the root together with the tree size.
\end{enumerate}

\cmd{nova verify} reports \code{signature\_ok}, \code{timestamp\_ok} and
\code{log\_integrity\_ok}, and, since v0.107, \code{timestamp\_trust} and
\code{sealer\_log}. Since v0.105 it also reports \code{identity\_trust} (\exper): one of
\code{self-asserted}, \code{local-ca-pinned}, \code{ca-anchored} or \code{none}. A
signature alone proves possession of a key. \code{identity\_trust} states how far that
key is bound to a real identity. \cmd{nova seal init} (ADR-0301) creates a local CA for
the \code{local-ca-pinned} level. Sealing remains opt-in.

\paragraph{Intact is not trusted.} Since v0.107, \code{timestamp\_trust} separates a
token's integrity from the trust placed in its time. It is \code{anchored} when the TSA
chain reaches an anchor the verifier supplied (\code{--tsa-ca-bundle}); \code{unanchored}
when the token is intact and bound to the envelope but no anchor was consulted, because
anyone can mint a TSA certificate and sign any time, so its \code{genTime} is a claim and
the output prints \code{UNANCHORED}; \code{absent}, since timestamping is opt-in; or
\code{invalid}. \code{timestamp\_ok} and the exit code keep their meaning. The Merkle
check reports where its result came from. On a host whose log holds other seals by the same
signer under the same root version, an entry missing from that log fails verification even
when the capsule carries a valid inclusion proof. A carried proof alone shows only that the
capsule's log metadata is self-consistent, because its tree head is neither signed nor
witnessed. The sealer's log is recognised by signer, not by the writer-controlled
\code{root\_version} in \code{log-entry.json}: a record that claims a newer version than the
log fails, and one that claims an older version is checked against the retired logs listed
under \code{previous\_merkle\_dbs} in \code{novaseal.yaml}, otherwise reported as
\code{sealer\_log: unverifiable-other-version}. \cmd{nova verify --strict} exits 1 on that
state, on an unanchored timestamp and on every other passing claim that rests on no anchor
(\code{unanchored\_claims}, for example a self-asserted signer).

Since v0.106 the signing intent (authored, reviewed, approved; FDA 21~CFR \S11.50) is bound
inside the signed payload (ADR-0309). An edited intent label fails verification, and a seal
made before v0.106 reports its intent as \code{label-only}, never as signed.

\paragraph{Further trust primitives} (all \exper):
\begin{itemize}
  \item a forward-secure key ratchet (\cmd{nova seal ratchet});
  \item Merkle consistency proofs, a Merkle Mountain Range and witness co-signing of
        tree heads;
  \item a hybrid, crypto-agile signature envelope;
  \item \code{did:key} identifiers with Verifiable Credentials;
  \item delegation (``acted-as'') chains and per-tenant key-encryption keys;
  \item an adversary-anchored ledger (\cmd{nova ledger anchor|verify|status}); without
        \code{--merkle-db}, a passing \cmd{nova ledger verify} shows only that a checkpoint's
        Merkle block is self-consistent, and exit 8 reports a checkpoint whose Merkle block does not verify.
\end{itemize}

\section{Merkle constructions and the price of the guarantee}
Three Merkle constructions coexist, and they are \textbf{not interchangeable}:
\begin{itemize}
  \item \fpath{evidence/merkle.py} is an RFC~6962-style~\cite{rfc6962,merkle} tree over the files of a capsule
        or evidence bundle. Since v0.106 each leaf commits to the file's relative path
        as well as its content (v2 leaves, ADR-0310), so renaming a file changes the
        root. Roots recorded earlier still verify, and v1 and v2 leaves are hashed under
        separate domain bytes, so accepting v1 cannot be used as a downgrade;
  \item \fpath{trust/novaseal/merkle.py} is the append-only signing log;
  \item \fpath{trust/novaseal/mmr.py} is a Merkle Mountain Range.
\end{itemize}
Mixing leaves or combiners across them silently yields a wrong root.

\begin{honest}[Verification cost: a withdrawn target (ADR-0268)]
Log verification recomputes the root from \emph{every} stored leaf hash. That full
recomputation \emph{is} the tamper detection for the stored hashes. The cost is linear
and measured. Computing the root alone takes 1361\,ms in-process for one million leaves.
End-to-end it takes about 1.9\,s on a CI runner with Postgres~16. The earlier target of
``p99 < 200\,ms at 1M entries'' was never met by any release. It has been
\withdrawn{} rather than deferred, and the release note that claimed it is struck
through, not deleted. A faster certificate-transparency-style path would give a weaker
guarantee, so adopting it requires a new ADR. Treat full verification as a scheduled
audit of roughly 1.2--2\,s per million entries, not as an interactive check.
\end{honest}

\begin{honest}[Reading a log root]
The signing-log root is versioned (ADR-0313). A log created from v0.106 on writes a
\emph{v2} root, the RFC~6962/9162 Merkle Tree Hash, which commits to the tree size. A log
created earlier stays \emph{v1} for its whole life. v1 pads an odd level by duplicating its
last node, so \code{[a,b,c]} and \code{[a,b,c,c]} share a root. Every root is therefore
carried with its \code{tree\_size} and \code{root\_version}: \cmd{nova verify} fails a log
entry that records a root without its tree size, and consistency proofs require both sizes. Releases before
v0.106 cannot verify the inclusion proof of a v2 seal, except at power-of-two tree sizes.
\end{honest}

\section{Evidence Bundles}
\cmd{nova export-evidence} packages one capsule, or since v0.102 a set of capsules,
into a ZIP for an auditor. The ZIP holds in-toto Statement v1 attestations, each signed
as a DSSE envelope, plus an RFC~3161 token when requested. The bundle schema carries a
\code{timestamp\_status} and an \code{admissibility\_status} (\code{self-authenticating},
\code{requires-foundation} or \code{incomplete}, ADR-0095). That status tells a reader
what the bundle can prove on its own and what still needs a witness.

Since v0.106, \cmd{nova verify <bundle>.zip} runs the bundle's whole verification recipe.
It recomputes \code{manifest\_hash} and every artifact digest, verifies the DSSE signature
of each attestation, and checks each statement's subject against the manifest. The run
statement's subject is also recomputed from the bundle's own \fpath{run-capsule/} copy, so
a capsule file edited after export fails even when every digest has been repaired. The
signing key ships inside the bundle, so a pass on its own proves internal consistency, not
identity: \code{--public-key} pins the expected signer, and without it the output says
the signer is not pinned. A bundle's RFC~3161 token is checked only when a
\code{--tsa-ca-bundle} anchor is given; otherwise the command says that its time is not
trusted.

Since v0.107 every member is signed. A members statement
(\fpath{attestations/bundle.\allowbreak{}intoto.\allowbreak{}json}) covers every file and the manifest's fields, so
rewriting \code{README.md}, the bundled schemas or \code{verifier\_instructions} fails
verification, \code{--public-key} included. A bundle built before it verifies as
``signed content only'' (\code{members\_signed: false}) and is never shown as fully
verified. A bundle's runs come only from its verified signatures, so a fabricated,
relabelled or stripped run, a duplicate ZIP member, or a run statement that signs no
\code{predicate.run\_id} fails. \code{POST /api/evidence/\{id\}/verify} in \cmd{nova serve}
runs this same verifier, where it once reported \code{valid: true} for tampered bundles.
Server-side signer pinning uses an operator trust list, \code{NOVAFABRIC\_EVIDENCE\_TRUSTED\_KEYS},
and reports \code{signer\_trust} as \code{anchored}, \code{unanchored} or \code{untrusted};
the command line pins with \code{--public-key}. Export never follows a symbolic link:
\cmd{nova export-evidence} refuses a capsule that contains one (exit 1, nothing written),
because the link's target, any file the exporting user can read, would otherwise be copied
into the signed bundle past redaction. The capsule Merkle root and in-toto subjects do not
follow links either.

\section{WORM storage, retention and legal holds}
Records can be made tamper-resistant at the storage layer as well as cryptographically.
WORM adapters cover S3 Object Lock in \code{COMPLIANCE} mode, Azure immutable blobs and
GCS Bucket Lock (\fpath{storage/}). The object capsule store adds S3-compatible
backends, including Ceph (\fpath{object\_capsule\_store/worm/}). A retention sweep that
respects WORM locks (\cmd{nova retention}) and legal holds (\cmd{nova hold
create|list|release}) complete the lifecycle. Since v0.107 the CLI and the API share one
legal-hold store (\fpath{registries/<r>/holds.jsonl} under \code{NOVAFABRIC\_HOME}). A hold
is registry-wide, so any active hold refuses \cmd{nova capsule delete} (\code{--force}
included) and the retention sweep, and stops PII erasure: the key is kept and
\cmd{nova pii erase} exits 3 (GDPR Art.~17(3)(e)). \cmd{nova retention apply} re-reads
holds and WORM locks before every action, and releasing a hold through
\cmd{nova serve} needs the \code{admin} scope.
The \code{nova-worm-conformance} package is a behavioural suite for storage targets.

\section{Two-person control}
Maker-checker is enforced in two places (Figure~\ref{fig:fig-09-maker-checker}), and a
third, protected labels, follows the registry's rules:
\begin{itemize}
  \item \textbf{Registry promotion (ADR-0058, ADR-0312).} \cmd{nova promote propose} and
        \cmd{nova promote approve} require \code{--identity}. Since v0.106 each signature
        is verified with the public key that an operator-held signer trust list
        (\code{NOVAFABRIC\_PROMOTE\_SIGNERS}, or \fpath{promote-signers.yaml} under
        \code{NOVAFABRIC\_HOME}) binds to that identity and its \code{proposer} or
        \code{approver} role. Separation of duties compares the \emph{bound} identities,
        and an approval is bound to one proposal. Only then does the transition pass the
        OPA policy gate and the eval-score gate. A direct promotion into
        \code{production} needs an explicit allow from the \code{promote\_direct} policy,
        which the shipped default denies. Since v0.107 every other route into
        \code{production} passes the same gate (403 without an allow; \code{force} does not
        bypass it): both servers' HTTP promote routes and \cmd{nova prompt promote};
        registration accepts only \code{development}; and \cmd{nova rollback} skips the
        gate only for a version the append-only status history shows was in production.
  \item \textbf{Capsule attestation (ADR-0059, ADR-0315).} The approval envelope binds the
        RFC~8785 canonical digest of the proposal envelope. \cmd{nova seal verify} runs five
        checks offline, each with its own exit code. It verifies the proposal envelope and
        its signer against policy, then the approval envelope and its signer. It checks
        that the bound digest matches, that the approver is not the proposer (exit 6), and
        that the timestamps are in order. Since v0.106 every proposal, approval, signed
        policy and bypass signer must chain to a pinned CA (\code{--ca-bundle} or
        \code{NOVAFABRIC\_PROMOTE\_CA\_BUNDLE}); with no trust root, verification refuses
        with exit 3. A bypass (\cmd{nova seal bypass}) counts only when it is signed by an
        approver in the current policy, for this capsule, and is unexpired by its
        \emph{signed} \code{valid\_until}. Since v0.107 it is also refused unless its
        \code{seal.bypass.create} audit entry is written, and a bypass made earlier stops
        counting until it is re-created.
  \item \textbf{Protected labels (ADR-0114, \exper).} \cmd{nova label propose-move} and
        \cmd{nova label approve-move} need an existing keyring key that the same trust list
        binds to the \code{proposer} or \code{approver} role. Before v0.107 approval minted a
        key for any unseen identity, so one person could satisfy both roles by passing a
        second \code{--identity}. Separation of duties now compares the bound identity and the
        key fingerprint, and only approvals that verify are counted. Relaxing a protection
        (\code{--unprotect}, fewer required approvals, a changed or removed
        \code{--policy-ref}) only proposes the change, and a trust-listed checker approves it
        with \cmd{nova label approve-move}; strengthening stays single-actor.
\end{itemize}

\begin{honest}[What two-person control proves]
Since v0.106 registry promotion and capsule attestation, and since v0.107 protected labels,
bind each signature to an identity the operator has vouched for. That binding is exactly as strong as the operator's trust list or pinned CA: whoever
controls them decides which key counts as which person. Login-level authentication of the
signer (OIDC in server mode, Sigstore) is not yet part of the check.
\begin{itemize}
  \item \code{--insecure-self-signed} (or \code{NOVAFABRIC\_PROMOTE\_INSECURE\_SELF\_SIGNED=1})
        accepts self-signed attestation signers for development. A regulated deployment
        must not set it.
  \item Proposals stored before v0.106 read as \code{unverified} and cannot be approved;
        a new \cmd{nova promote propose} supersedes them.
\end{itemize}
\end{honest}

\nffigure{fig-09-maker-checker}{Maker-checker. The order is as built: the separation-of-duties
check, then the policy gate, then the transition. Every outcome is audit-logged.}

Every policy decision, promotion, approval, hold and deletion is appended to a SHA-256
hash-chained audit log. Since v0.107 concurrent appenders are serialised with a file
lock, so the chain cannot fork. The log exports as JSONL, OCSF or CEF for a SIEM
(\cmd{nova audit-log export}, ADR-0191).

\chapter{Lineage and the knowledge graph}\label{ch:lineage}

\section{Event-derived, rebuildable lineage}
Lineage is not a separate system of record. Capture writes typed edges into each
capsule's \code{lineage.jsonl}, and a graph store indexes them. The store can always be
rebuilt from the capsules. Every edge carries a unique ULID identifier, and importing a
capsule replaces the edges the store holds for that run instead of adding to them, so
importing the same capsule twice leaves the same graph
(Figure~\ref{fig:fig-10-lineage-model}). Since v0.107 the importer and the writer read and
rewrite a capsule's lineage and source streams through the no-follow layer
(Chapter~\ref{ch:capture}): a symbolic link or special file at \code{lineage.jsonl},
\code{assets.jsonl} or \code{capsule.yaml} is an error rather than an empty result, so a
lineage index never reports zero edges over a record it could not read.

\nffigure{fig-10-lineage-model}{The lineage vocabulary: capture-time edges (top) and the
distributed, session and memory vocabulary (ADR-0044, ADR-0122, ADR-0143).}

\cmd{nova lineage} answers the questions an incident raises:
\begin{itemize}
  \item \code{provenance}: what led to this run?
  \item \code{blast-radius}: what is downstream of this asset?
  \item \code{replay-chain}: which runs replay which?
  \item \code{time-travel}: what did the graph look like at time \emph{t}?
  \item \code{root-cause}: which upstream change explains this failure?
\end{itemize}
The memory edges serve one specific purpose: when a memory item is poisoned, they
distinguish \emph{who wrote} the bad item from \emph{who read} it.

\section{Tiered backends behind one interface}
\begin{nftable}
\begin{tabularx}{\linewidth}{@{}lLl@{}}
\toprule
\thead{Backend} & \thead{Role} & \thead{Status} \\
\midrule
SQLite (recursive BFS) & local default, below about 1M edges & \works \\
Postgres \code{WITH RECURSIVE} & server-mode tier (v0.68) & \exper \\
KuzuDB (embedded) & the 1M--1B edge production candidate; bulk \code{COPY} write path (v0.94) & \exper \\
Apache AGE (openCypher) & Postgres-native graph tier (v0.69) & \exper \\
JanusGraph (Gremlin) & distributed billion-edge tier (v0.70) & \exper \\
\bottomrule
\end{tabularx}
\end{nftable}

All four at-scale backends are implemented, and testcontainers verify them against
live servers. No \code{NotImplementedError} stubs remain in \fpath{lineage/backends/}.
Since v0.102 one shared behavioural contract test runs against every backend. That
matters: JanusGraph's first live run exposed two real bugs that unit tests against mocks
had missed. In the cluster tier, \cmd{nova lineage consume} derives \code{SPAWNED\_BY}
edges from canonical \code{RunStarted} events on NATS that carry a \code{parent\_run\_id}.
Deduplication by \code{event\_id} happens here, downstream of the collector. Since v0.95 (ADR-0220) the producer-to-consumer path is verified end-to-end. A test
runs a real parent and child capture, feeds the resulting spool segments to the consumer
and asserts the edge. The NATS JetStream transport itself is covered by mocked tests.

\section{Interoperability}
AI lineage lands in existing data catalogues instead of a new silo:
\begin{itemize}
  \item \cmd{nova lineage emit-openlineage} emits OpenLineage RunEvents;
  \item \cmd{nova lineage export-prov} exports W3C PROV-JSON or PROV-N;
  \item fine-grained row and transformation facets (v0.77) carry column-level detail.
\end{itemize}

\section{The capsule knowledge graph}
\exper{} A separate KuzuDB knowledge graph (\fpath{kg/}, ADR-0067) aggregates entities
\emph{across} runs: which agents call which models, which tools co-occur, which MCP
servers provide them. It includes an alias table and a human review queue for entity
resolution (\cmd{nova kg alias}, \cmd{nova kg entity-queue}). Phase~1 of the Security
Provenance Knowledge Graph (ADR-0111) adds \cmd{nova kg detect}, \cmd{nova kg attack-path} and
\cmd{nova kg blast-radius}, which reason over security-relevant paths through tools,
credentials and data. Hybrid hunting, temporal validity and an RDF store are \planned.
Since v0.107 the knowledge-graph consumer never follows a link: a linked directory or ZIP
is not a candidate and a linked event stream is not ingested. In \cmd{nova serve}, the
\code{capsule\_path} of the KG-ingest and lineage-import routes is a bare run id or a
path inside an allowlist of roots (\code{\$NOVAFABRIC\_HOME}, the capsule directory, and any
added through \code{NOVAFABRIC\_SERVE\_PATH\_ROOTS}); \code{..} segments, symlink escapes and
the server's own secrets, such as its token stores, are refused (403).

\chapter{Governance, compliance and assurance}\label{ch:governance}

The governing principle: \textbf{humans approve policies, risk boundaries and exceptional
actions; machine-enforced policy governs and records everything else.} Every decision
leaves evidence (Figure~\ref{fig:fig-11-policy-audit}).

\nffigure[0.62\linewidth]{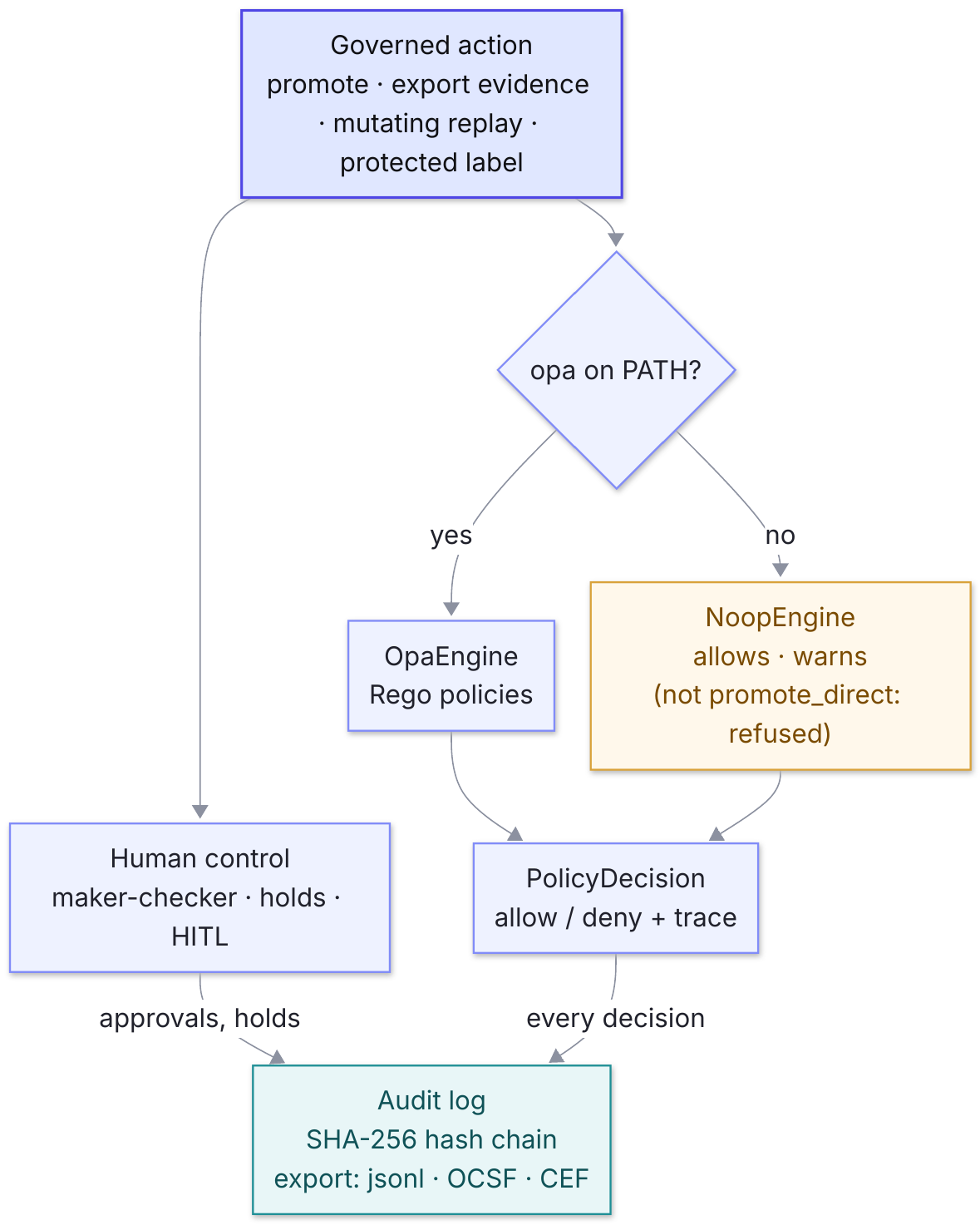}{Governed actions. A policy decision and any
human control both land in the hash-chained audit log, which exports to a SIEM.}

\section{Policy engine}
Promotion, evidence export, replay with any write-permitting flag and
protected labels are gated by OPA/Rego policies (\fpath{policy/}, default policies in
\fpath{policies/novafabric/defaults/}). Decisions are explainable
(\cmd{nova policy explain}), testable (\cmd{nova policy test}) and can be signed
(\cmd{nova policy sign}). OPA runs as an external \code{opa} binary with a 30-second
timeout. NovaFabric embeds no policy runtime, which keeps the dependency licence tier
clean and the policy language standard. Since v0.107 every \code{opa} process
NovaFabric starts inherits no environment and runs with capabilities that remove
\code{opa.runtime}, \code{http.send} and \code{net.lookup\_ip\_addr}, so Rego cannot read the
server's secrets or reach the network. If the restricted capabilities cannot be derived
from the installed binary, nothing is evaluated and the caller is denied
(\fpath{policy/\_opa\_sandbox.py}). Under \cmd{nova serve}, caller-supplied Rego
(\code{policy\_source}, or \code{bundle\_path} on the test route) needs the \code{admin}
scope.

\paragraph{Every route into production is gated.} \works{} A new asset enters the
registry as \code{development}: \cmd{nova register}, \code{POST /api/assets} and
\code{POST /v0/assets} refuse a spec that declares any other \code{status:}
(HTTP~422 on both routes, with the error code \code{registration\_status\_not\_allowed}
on \cmd{nova server}; exit~1 on the CLI). Every other move into
\code{production} then goes through the \code{promote\_direct} policy action or
maker-checker: \cmd{nova promote direct}, \cmd{nova rollback}, \cmd{nova prompt promote},
both \cmd{nova serve} promote routes and \code{PUT /v0/assets/\{id\}/promote} on
\cmd{nova server}. A refusal is HTTP~403, and \code{force} does not bypass it. Rollback
restores without a policy decision only a version that the append-only
\code{asset\_status\_history} table shows was in production, bound to the asset id and the
spec's SHA-256. A version that was ever in production can never be registered again.

\paragraph{Protected labels and legal holds.} Moving a protected label is a
maker-checker step (\cmd{nova label propose-move}, \cmd{nova label approve-move}; \exper). Both
roles need an existing keyring key bound to them in the signer trust list
(\cmd{nova promote keygen}), and separation of duties compares the bound identity and the
key fingerprint, so a second \code{--identity} cannot make one person both. Relaxing a
protection (\code{--unprotect}, fewer approvals, a changed policy reference) is itself only
a proposal that a trust-listed checker must approve. \works{} Legal holds live in one store,
\code{registries/<registry>/holds.jsonl}, for the CLI and the API. A hold blocks capsule
deletion (\code{--force} included), PII erasure and the retention sweep, and
\cmd{nova retention apply} re-reads holds and WORM locks before every action. Placing a
hold needs \code{operate} under \cmd{nova serve}; releasing one needs \code{admin}.

\begin{honest}[Where policy gates stop]
If \code{opa} is not on \code{PATH}, \code{get\_policy\_engine()} warns and returns a
\code{NoopEngine} that \emph{allows} every request (\fpath{policy/\_\_init\_\_.py}).
Local mode stays usable on a machine without OPA, but policy gates are then off. The
one deliberate exception is \code{promote\_direct}: a no-op ``allow'' is not an explicit
allow, so single-actor promotion into production is refused until OPA is installed
(\fpath{registry/service.py}). A regulated deployment must install OPA and treat that warning as a configuration error.
\end{honest}

\section{Human oversight as evidence}
\exper{} The \code{hitl} package (ADR-0150) records human--agent accountability as
first-class evidence, with each record handled by its own commands:
\begin{itemize}
  \item conversation provenance, decision context, overrides and rationale
        (\cmd{nova hitl});
  \item consent receipts and their withdrawal (\cmd{nova consent});
  \item hand-offs and ``acted-as'' delegation.
\end{itemize}
All of it is record-only: NovaFabric documents who decided what, and does not decide
for them. A five-level approval taxonomy (L0 to L4) and a human-impact ledger are
\fdesign; today only \code{HumanApprovalEvent} exists in code.

\section{The accountability spine}
\exper{} Four mechanisms, shipped in v0.55--0.56, bind a run to claims an outside party
can check:
\begin{itemize}
  \item \textbf{Energy-anchored receipts} (\cmd{nova energy probe|attest|verify|report}):
        RAPL sampling and Slurm \code{sacct} energy, attached as evidence.
  \item \textbf{Adversary-anchored ledger} (\cmd{nova ledger}).
  \item \textbf{Replay attestation} (\cmd{nova evidence attest-replay}): since v0.105 a
        diverged mocked replay attests \code{mismatch}. Since v0.106 \code{--mode exact}
        refuses, and signs nothing, for a run that is not exact-eligible, because exact
        mode re-executes nothing.
  \item \textbf{Safety-case compiler} (\cmd{nova safety-case build|export|verify},
        ADR-0095): Claims--Arguments--Evidence cases from templates for a generic
        frontier case, EU AI Act Annex~IV and NIST AI RMF.
\end{itemize}

\begin{honest}[Ledger anchoring]
The only anchor the ledger writes today is a local finalize anchor
(\code{method: "none"}), which the verifier accepts. External anchoring to a TSA or a
transparency log is not yet in the ledger package. Until it is, the ledger proves
internal consistency, not independent witness. Likewise, without
\code{--merkle-db}, \cmd{nova ledger verify} exit~0 shows only that a checkpoint's Merkle
block is self-consistent; with it, the supplied log must hold the checkpoint.
\end{honest}

\section{Compliance machinery}
\nfExportCmds{} \cmd{nova export-*} commands turn capsules into the artefacts that
regulatory and archival processes consume. All regulatory exporters are \exper; the core
evidence bundle is \works. Since v0.107 no exporter follows a symbolic link: a capsule that
contains one is refused and nothing is written, and a required capsule file that is a link,
FIFO, directory or oversize is an error, never empty output.

\begin{nftable}
\begin{tabularx}{\linewidth}{@{}P{0.23\linewidth}L@{}}
\toprule
\thead{Area} & \thead{Exporters (\cmd{nova export-}\dots)} \\
\midrule
EU AI Act & \code{annex-iv}, \code{public-annex-viii}, \code{compliance} (ISO~42001, NIST GenAI profile, GPAI Art.~53, post-market monitoring) \\
Risk \& governance & \code{nist-rmf}, \code{model-risk} (SR~26-2 / SR~11-7), \code{model-independence}, \code{control-attestation}, \code{rai-scorecard}, \code{system-card} \\
Privacy \& security & \code{ropa} (GDPR Art.~30), \code{hipaa-proof}, \code{nis2}, \code{whistleblower} \\
Finance \& regulated records & \code{adverse-action} (ECOA / Reg~B), \code{cat}, \code{part11} (21~CFR~Part~11), \code{retention} \\
Public sector & \code{public-disclosure}, \code{public-incident}, \code{transparency-register}, \code{foia}, \code{citizen-explanation}, \code{election-disclosure}, \code{accessibility-claim} \\
Provenance \& archival & \code{evidence}, \code{aibom} (CycloneDX~1.7), \code{c2pa}, \code{rocrate}, \code{rocrate-science}, \code{examiner} (BagIt, FDA PCCP, ISO~42001), \code{blob} \\
\bottomrule
\end{tabularx}
\end{nftable}

\cmd{nova audit} scores evidence coverage against \nfAuditProfiles{} built-in profiles:
EU AI Act high-risk, GDPR, ISO~42001, NIST AI RMF (the default), scientific
reproducibility and SOC~2 Type~2.

\paragraph{GDPR Art.~17 crypto-shredding.} \exper{} \cmd{nova pii erase} destroys a
data subject's AES-GCM data-encryption key and issues an erasure receipt (ADR-0069). The
design makes ciphertext unreadable without touching the immutable capsule. Since v0.106,
\cmd{nova capture --pii-subject <id>} encrypts detected PII and every literal occurrence
of the subject id under that subject's key at capture time (ADR-0314, requires
\code{NOVA\_PII\_PEPPER}). After erasure those tokens cannot be opened and readers render
\code{[erased]}. Erasure makes PII unreadable \emph{only} in capsules captured that way;
binary or oversize artifacts, media blobs and in-process (adapter or decorator) capsules
are not covered, and they still need the masker, capture levels and retention to meet an
erasure obligation. If the key is unavailable at capture, the PII is masked and never
stored in plaintext. A legal hold wins over erasure (GDPR Art.~17(3)(e)): while any hold
is active in any registry, \cmd{nova pii erase} keeps the key, writes a held receipt and
exits 3 (the REST route answers HTTP~409). Holds are registry-wide today, so one active
hold defers every erasure until it is released. The
\code{nova erasure} group used to report success without doing anything. It now fails
loudly and points to the working path, a deliberate correction recorded in its module
docstring.

\paragraph{Supply chain and MCP.} \cmd{nova mcp} scans MCP servers for supply-chain risk.
\cmd{nova dataset} and \cmd{nova toolschema} record signed dataset provenance cards and
tool-schema evidence. \cmd{nova assure} checks evidence against the OWASP LLM Top-10
(2025).

\section{Honest numbers: cost and evals}
\paragraph{Absent is not zero (ADR-0234).} An unpriced model used to report \$0.00, which
reads as ``free''. Cost aggregates now track \code{unpriced\_calls} and
\code{unpriced\_models}. An aggregate that cannot be computed honestly refuses, with a
reason and a remedy, instead of rendering a number. \cmd{nova cost estimate} reports
unpriced models in their own \code{basis: "unpriced"} row, kept out of the totals.

\paragraph{Evals are promotion gates, not scores.} Eval suites (MMLU, TruthfulQA, GAIA,
SWE-bench, AgentBench, smoke) feed the Rego regression gate on promotion. Since
v0.106 (ADR-0308), container suites \emph{require} a user-supplied
image and digest. No NovaFabric eval image is published, and an unset variable now fails
before Docker is touched, naming exactly what is missing.

\chapter{Product surfaces}\label{ch:surfaces}

\section{The CLI is canonical}
\works{} The CLI is the reference interface. Every other surface either calls the same
library code or shows the equivalent command. At v\nfVersion{} it registers \nfTopCmds{}
top-level commands and \nfCmdPaths{} runnable command paths; Appendix~\ref{app:cli}
lists them all. Two guard tests keep the documentation in step. One requires every
top-level command to have a section in \fpath{docs/cli-reference.md}. The other forbids
a documented heading for a command that no longer exists. The story still fits in six
lines:

\begin{shell}
nova capture python my_agent.py        # record a run as a capsule
nova validate  <run-id>                # check the capsule schema
nova replay    <run-id> --mode mocked  # re-run under a stated contract
nova diff <run-a> <run-b> --assert-no-regressions   # CI gate
nova verify    <run-id>                # signature, time, log, trust
nova export-evidence <run-id>          # signed bundle for an auditor
\end{shell}

\section{The local dashboard}
\exper{} \cmd{nova serve} starts a local FastAPI server with a single-page dashboard
(\fpath{ui/dashboard/}). It reads the same capsules and indexes as the CLI. Its
\nfDashTabs{} tabs fall into \nfDashGroups{} journeys:

\begin{nftable}
\begin{tabularx}{\linewidth}{@{}P{0.22\linewidth}L@{}}
\toprule
\thead{Journey} & \thead{Tabs} \\
\midrule
Overview & Home, Analytics, Cost, Dashboards \\
Runs \& debug & Runs, Diff, Capture \\
Govern \& promote & Registry, Governance, Eval, Risk, Policy, Holds \\
Provenance \& trust & Lineage, KG, Evidence, Seal, Spine, Audit \\
Compliance & Compliance, Incidents, Schema \\
Platform & Infra, Storage, Ops, Alerts, Admin, Commands \\
Reports \& export & Reports, Export \\
\bottomrule
\end{tabularx}
\end{nftable}

The Commands tab is a CLI command builder, and each tab shows the CLI equivalent of its
actions. Dashboards are portable, versioned JSON files (\cmd{nova dashboard}, ADR-0235).
They can be diffed, reviewed and committed like code. Optional live topology views,
behind \code{--topology} and \code{--tv5}, stream fleet graphs over WebSockets.

\section{Authorization on every route (ADR-0228)}
\nfServeRoutes{} HTTP routes are mounted by default. Each is classified in one
declarative \code{(method, path)} table, \code{ROUTE\_SCOPES}, with \nfAuthzRoutes{}
entries covering the optional topology and TV-5 routes too
(Figure~\ref{fig:fig-12-serve-authz}). Scopes nest: \code{admin} includes
\code{operate}, which includes \code{read}. \code{audit} is orthogonal: it grants audit and
read access and nothing else.

\begin{keyidea}[Deny by default]
A route missing from the table is denied to every caller, an admin and an anonymous
request included (since v0.107; before, an anonymous request fell through to the route's
own authentication, which an unclassified route might not have). Denials are audited. On a
classified route, an unauthenticated request still reaches that route's own authentication
dependency, which answers 401. A completeness guard test fails the build if any mounted
route is unclassified, so a new endpoint cannot ship ungoverned by accident.
\end{keyidea}

A stored token scope that is not recognised (a typo, another case, padding) now grants
nothing; before v0.107 it fell back to \code{admin}. Only a record with no scope field at
all is still treated as legacy \code{admin} (ADR-0228 D4). WebSocket routes are outside the
table and keep their inline Host and token check.

\nffigure[0.5\linewidth]{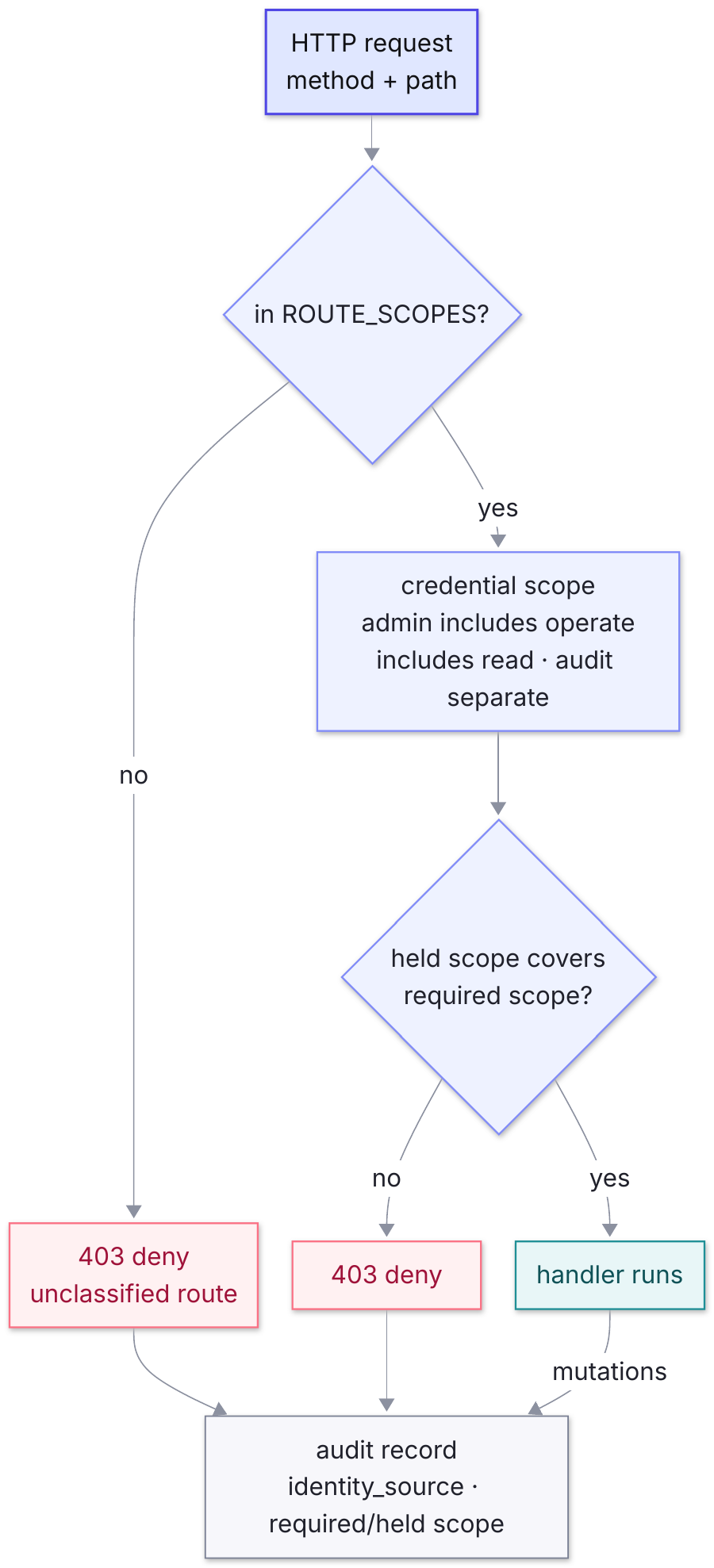}{Request authorization in \cmd{nova serve}.}

The neighbouring ADRs harden the same surface:
\begin{itemize}
  \item \textbf{ADR-0229, tenancy posture.} Every store declares itself tenant-aware,
        tenant-agnostic or unsafe, and the declaration is checked against the real
        schema. Multi-tenant mode is \emph{refused at startup} while any store on a read
        path is unsafe. This is stricter than the ADR's original plan of returning 503
        per endpoint, because a partly scoped API that looks healthy is worse than one
        that refuses to start.
  \item \textbf{ADR-0230, secure-by-default chart.} The Helm chart defaults to
        authenticated server mode. Rendering it with \code{insecure: true} fails unless
        the operator writes a literal acknowledgement string.
  \item \textbf{ADR-0231, audit enrichment.} Records carry
        \code{actor.identity\_source} (\code{shared-token}, \code{credential} or
        \code{federated}). A shared token is no longer exported to a SIEM as if it were
        a named user.
\end{itemize}

\paragraph{What the server refuses to be used for (v0.107).} Four boundaries close ways
for a low-privilege token to reach the host.
\begin{itemize}
  \item \textbf{Caller-supplied Rego needs \code{admin}.} \code{POST /api/policy/check}
        with \code{policy\_source}, and \code{bundle\_path} on \code{/api/policy/test},
        return 403 to a lesser credential. Every \code{opa} runs with no inherited
        environment and without \code{opa.runtime}, \code{http.send} and
        \code{net.lookup\_ip\_addr} (Chapter~\ref{ch:governance}). Before v0.107 a
        \code{read} token could read the server token through Rego and become \code{admin}.
        \code{GET /api/policy/explain} is \code{audit}-scoped with exact matching.
  \item \textbf{Request-body paths are allowlisted.} A path in a request body (evidence
        export, capsule migrate, policy sign and test, KG and lineage \code{capsule\_path},
        metadata rebuild) is resolved and must lie inside a root. Writes go only to
        \code{\$NOVAFABRIC\_HOME/exports}, \code{\$NOVAFABRIC\_HOME/evidence} and
        \code{\$NOVAFABRIC\_EVIDENCE\_DIR}; reads may also use \code{\$NOVAFABRIC\_HOME}
        and the capsule directory. Token stores, \code{dek.db}, ratchet state, the audit
        logs and (for source reads) key material are refused even inside a root. A \code{..}
        segment or a symlink escape is a 403.
        \code{NOVAFABRIC\_SERVE\_PATH\_ROOTS} adds roots; the old
        \code{NOVAFABRIC\_SERVE\_ALLOW\_ANY\_PATH} switch is no longer honoured.
  \item \textbf{The token never reaches a log.} The audit log, the SIEM export and
        \code{GET /api/admin/tokens} carry a 12-hex-character SHA-256 fingerprint, not the
        token's first eight characters; earlier records keep the prefix, as the chain is not
        rewritten.
  \item \textbf{Capsules are not followed through links.} No \cmd{nova serve} route
        follows a symbolic link in a capsule; hard links remain a documented limit.
\end{itemize}

\section{Server mode}
\exper{} \cmd{nova server start} runs the multi-user \code{/v0} API, \nfServerRoutes{}
routes in its default configuration, specified in \fpath{api/openapi.yaml}. It provides:
\begin{itemize}
  \item OIDC bearer validation against a JWKS cache (asymmetric algorithms only) and RBAC
        with roles \code{reader} < \code{writer} < \code{admin} plus a separate \code{auditor};
  \item Ed25519 offline tokens for air-gapped and Slurm sites (ADR-0018);
  \item hashed API keys, SAML~2.0 SP, SCIM~2.0, step-up authentication, webhooks, usage
        metering, quotas and rate limiting.
\end{itemize}
Since v0.107 \code{POST /v0/evidence} takes no filesystem path from the client: it
refuses \code{output\_path} (400) and writes only \code{<evidence dir>/<run\_id>.zip}, and
every id that becomes a file name, from a request, manifest, event or \code{capsule.yaml},
must be one path segment. A symlinked store entry is refused. Moving an asset into
\code{production} through \code{PUT /v0/assets/\{id\}/promote} needs a
\code{promote\_direct} allow (Chapter~\ref{ch:governance}).

\cmd{nova login} is demo-only: it runs the RFC~8628 device flow against demo endpoints that
exist only when the server sets \code{NOVAFABRIC\_SERVER\_DEMO\_DEVICE\_GRANT=1} (off by
default), never contacts an identity provider, and stores a token the server's API does not
accept. Against a real server use an IdP-issued bearer token, an API key or
\cmd{nova server issue-token}.

\section{SDKs, collector and integrations}
\begin{itemize}
  \item \textbf{Python:} the in-process SDK and \code{record} façade, plus a typed client
        for the \code{/v0} API (\fpath{client/}, \exper).
  \item \textbf{TypeScript:} \code{@novafabric/sdk}, with types generated from the
        OpenAPI contract and no runtime dependencies (\exper).
  \item \textbf{Go collector} (\fpath{collector/}): four binaries: collector, HPC hub,
        spool forwarder and verifier (Chapter~\ref{ch:scale}).
  \item \textbf{Claude Code plugin} (\fpath{integrations/claude-plugin}): skills that
        instrument an agent project and deploy NovaFabric.
\end{itemize}

\section{Distribution and supply chain}
\begin{itemize}
  \item \textbf{PyPI:} \code{novafabric} (Apache-2.0, Python ≥ 3.12), published by OIDC
        Trusted Publishing with SLSA provenance and a CycloneDX SBOM.
  \item \textbf{Container image:} \code{ghcr.io/mskazemi/novafabric} (amd64, arm64),
        keyless-cosign signed, with SBOM and provenance, Trivy-scanned at release; the Helm chart
        goes to an OCI registry with the version from the release tag.
\end{itemize}

\chapter{Architecture at scale}\label{ch:scale}

\begin{pullquote}Light at the edge. Heavy in the service plane. Distributed by child
capsules. Scalable by aggregation.\end{pullquote}

\begin{figure}[!t]\centering
\begin{minipage}[c]{0.36\linewidth}\centering
\includegraphics[width=\linewidth,height=0.58\textheight,keepaspectratio]{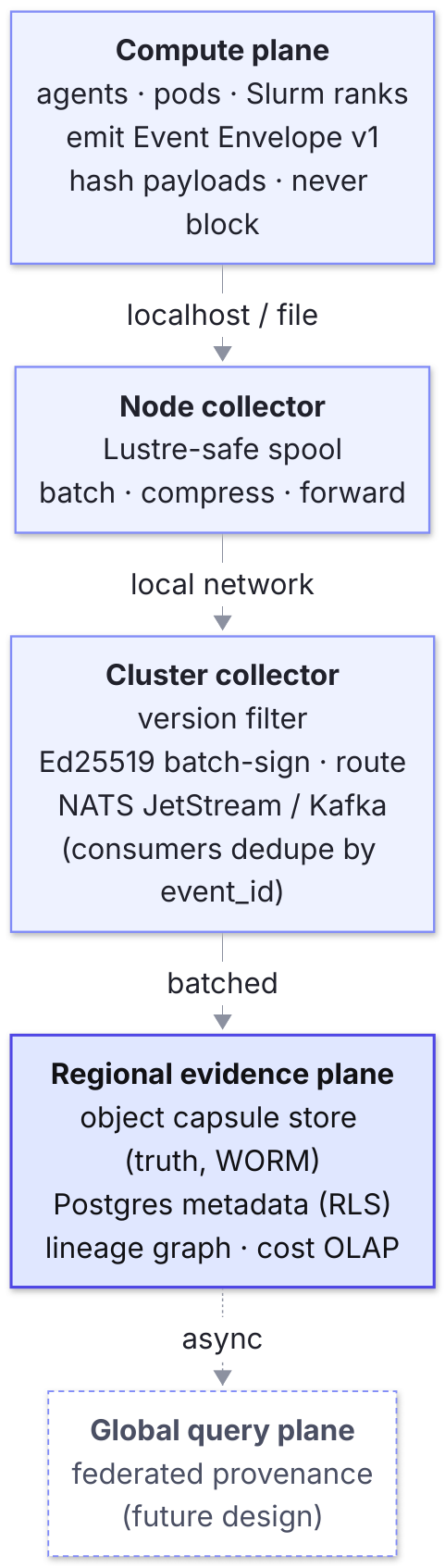}
\end{minipage}\hfill
\begin{minipage}[c]{0.58\linewidth}\small
\textbf{Compute plane.} Agents, pods and Slurm ranks emit Event Envelope v1 to a local
spool, hash their payloads and never block.\par\smallskip
\textbf{Node collector.} A Lustre-safe spool batches, compresses and forwards.\par\smallskip
\textbf{Cluster collector.} Drops envelopes of an unknown version, batch-signs with
Ed25519 and routes over NATS JetStream (primary) or Kafka. Delivery is at-least-once, and
consumers deduplicate by \code{event\_id}.\par\smallskip
\textbf{Regional evidence plane.} The object capsule store holds the truth under WORM.
A Postgres metadata index runs under row-level security, alongside the lineage graph
and cost analytics.\par\smallskip
\textbf{Global query plane.} Federated provenance across regions, \fdesign.
\end{minipage}
\caption{Deployment planes, from compute node to global query.}\label{fig:fig-13-cluster-planes}
\end{figure}

All cluster-scale phases, 0 to 6, are implemented and shipped as \exper. Compute nodes
do computation. On the cluster path a compute node writes Event Envelope v1 records, each
with its payload and the SHA-256 of that payload, to a local spool and never blocks;
signing, deduplication, routing and indexing happen on dedicated service nodes
(Figure~\ref{fig:fig-13-cluster-planes}). The envelope does not decide what a payload
contains: that is decided at capture (Chapter~\ref{ch:capture}), where prompts and
responses are stored at every capture level today.

\section{The Go collector}
\exper{} The collector (\fpath{collector/}) implements a demanding HPC constraint: a
\textbf{Lustre-safe spool}. On a parallel filesystem, rename is the only atomic
primitive, so the spool uses no \code{flock} and no \code{mmap}. It syncs each segment
to disk (\code{fsync}) before the rename commit, evicts FIFO, and returns back-pressure as
an error instead of blocking the agent. A custom OpenTelemetry Collector processor
canonically encodes each batch and signs it with Ed25519. It refuses to forward an
unsigned batch when the key service is unreachable. In the HPC profile, a Slurm Prolog
starts a per-job NATS leaf and signing happens at the cluster hub. Compute nodes
therefore have no key-service dependency, and air gaps hold. Where the Go binary is not
deployed, a small Python receiver (\code{collector\_app}, \code{POST /capsule}) feeds
capsule events into the evidence-fabric accumulator. It has no spool and no signing, so it
is a lightweight ingest path, not a replacement for the collector. Since v0.107 it
authenticates. With \code{NOVA\_COLLECTOR\_TOKEN} set, every \code{POST /capsule} must carry
the token in \code{X-Nova-Token}, from any address, and the comparison is constant-time.
Without a token it serves only a loopback caller with a \code{localhost} \code{Host} header
and no proxy header, and logs a warning at startup. A same-host reverse proxy that adds no
forwarding header is indistinguishable from a local caller, so the token must be set
whenever anything proxies to it. The tenant belongs to the authenticated caller
(\code{NOVA\_COLLECTOR\_TENANT}, default \code{default}); a body or event that names another
tenant is refused with 403, so one collector and one token serve one tenant, and a
malformed event is refused with 422 and stores nothing from its batch.

\paragraph{Batch-scheduler runners.} \exper{} The Slurm and LSF runners put the job script on
the submit command's standard input (\cmd{sbatch}, \cmd{bsub}), so it appears in no process
argument list and in no file in the shared capsule directory. Slurm, LSF and PBS quote every
argument with \code{shlex.quote}: a \code{\$}, backtick, \code{;} or \code{\&} in a
command argument or an \code{extra\_env} value reaches the job literally and is neither
expanded nor executed, as with the local runner. The Slurm and LSF submit commands start
from a minimal environment, not the user's shell (ADR-0311).

\section{Distributed runs and the fail-open arrival policy}
\nffigure[0.62\linewidth]{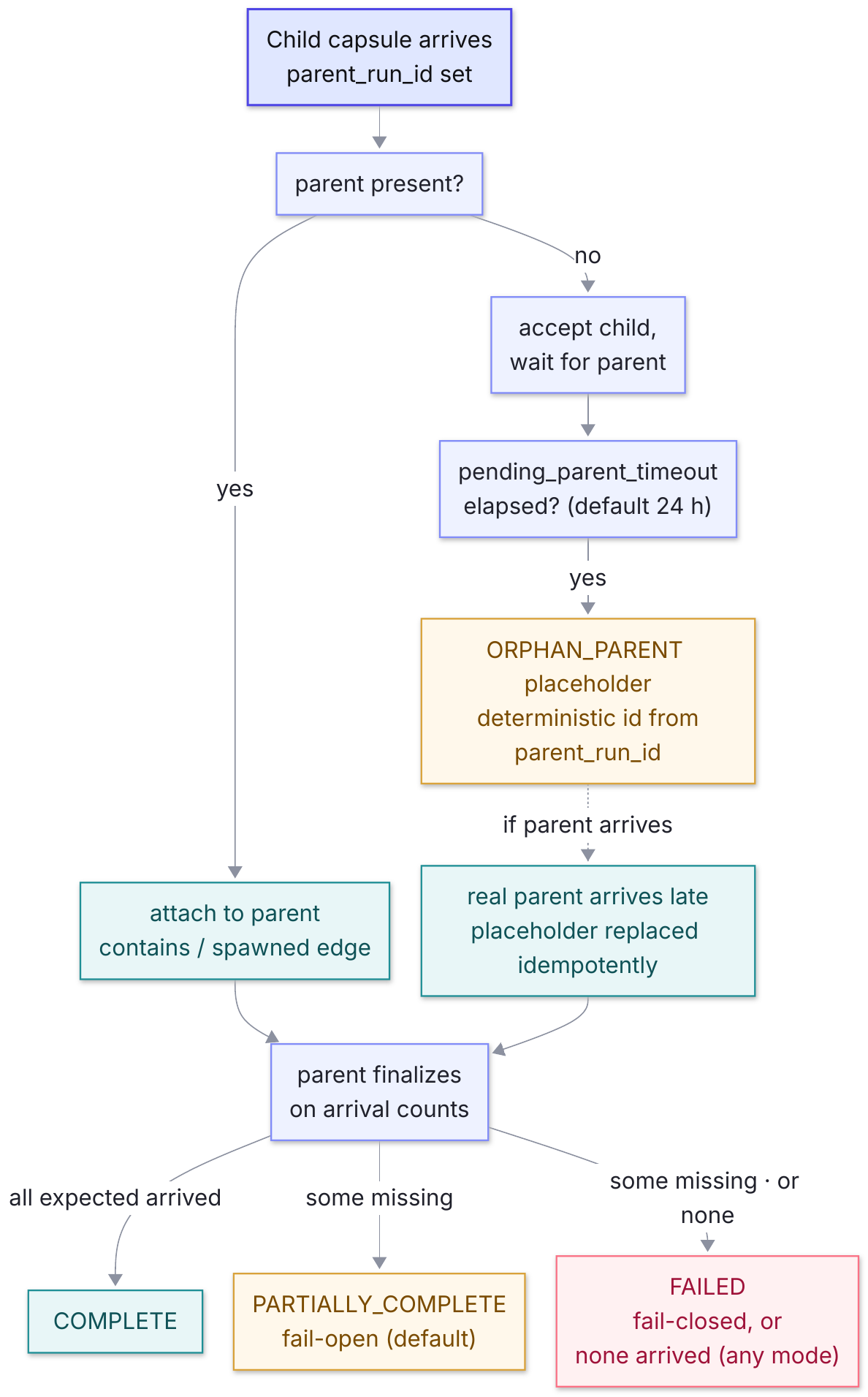}{The designed child-arrival, orphan and parent-finalization model (\fpath{capsule/orphan.py}, \fpath{capsule/writer.py}, \fpath{capsule/validator.py}); see the box below for what is wired.}

\exper{} A distributed job must not lose 999 workers' evidence because its driver
crashed. The parent/child capsule model is therefore designed to accept every valid child
capsule, even when its parent never arrives (Figure~\ref{fig:fig-14-parent-child}). After the
configurable \code{pending\_parent\_timeout} (default 24\,h), an orphan child is linked
to a synthetic \code{ORPHAN\_PARENT} placeholder. The placeholder's identifier is derived
deterministically from the missing \code{parent\_run\_id}, so every orphan of the same
parent resolves to the same placeholder. If the real parent arrives late, it replaces
the placeholder idempotently. A parent finalizes on \emph{arrival counts}. If every
expected child arrived, it is \code{COMPLETE}; if none arrived, \code{FAILED}. If only
some arrived, it is \code{PARTIALLY\_COMPLETE} under the default \code{fail-open} mode and
\code{FAILED} under \code{fail-closed}. A child's own \code{FAILED} status does not change
this outcome; it is recorded on the child. When the parent declares no expected count, it
is \code{COMPLETE} once any child has arrived.

\begin{honest}[What is wired today]
This state machine (\code{ParentCapsuleTracker}, \code{OrphanManager}) is tested library
code. No runtime path calls it. \cmd{nova capture} does not count arrivals, finalize a
parent or write an \code{ORPHAN\_PARENT} placeholder, and the \code{CapsuleWriter} that
writes the parent manifest is not the one \cmd{nova capture} uses. What does run:
\cmd{nova run show --with-children} assembles the child tree and lists children whose
parent is absent, under a synthetic root that exists in memory only. \cmd{nova run validate-distributed}
(and \code{POST /api/runs/\{id\}/validate-distributed}) reads a parent manifest's recorded
status and maps it to exit 0, 2 or 1. \cmd{nova server} refuses a child upload whose parent
is absent (409, \code{parent\_not\_found}) until the child is 24\,h old. A parent therefore
reaches a terminal status only if something outside \cmd{nova capture} finalizes it.
\end{honest}

\section{Polyglot persistence with one source of truth}
\exper{} Each backend serves the workload it fits best (ADR-0022), and one rule holds:
the object capsule store is the truth, and everything else is rebuildable
(Figure~\ref{fig:fig-15-storage-stack}).
\begin{itemize}
  \item \code{put\_capsule} is a five-step WORM protocol with optional zstd dictionary
        compression.
  \item \code{get\_capsule} pins SHA-256 and walks an Iceberg-style manifest chain by
        default, raising \code{ChainIntegrityError} on a broken link.
  \item The Postgres metadata store enforces tenant isolation by \code{SET LOCAL} at the
        start of every transaction, under a dedicated non-superuser application role.
        Its security gate tests run as that role, because a superuser bypasses
        row-level security.
\end{itemize}

\nffigure[0.86\linewidth]{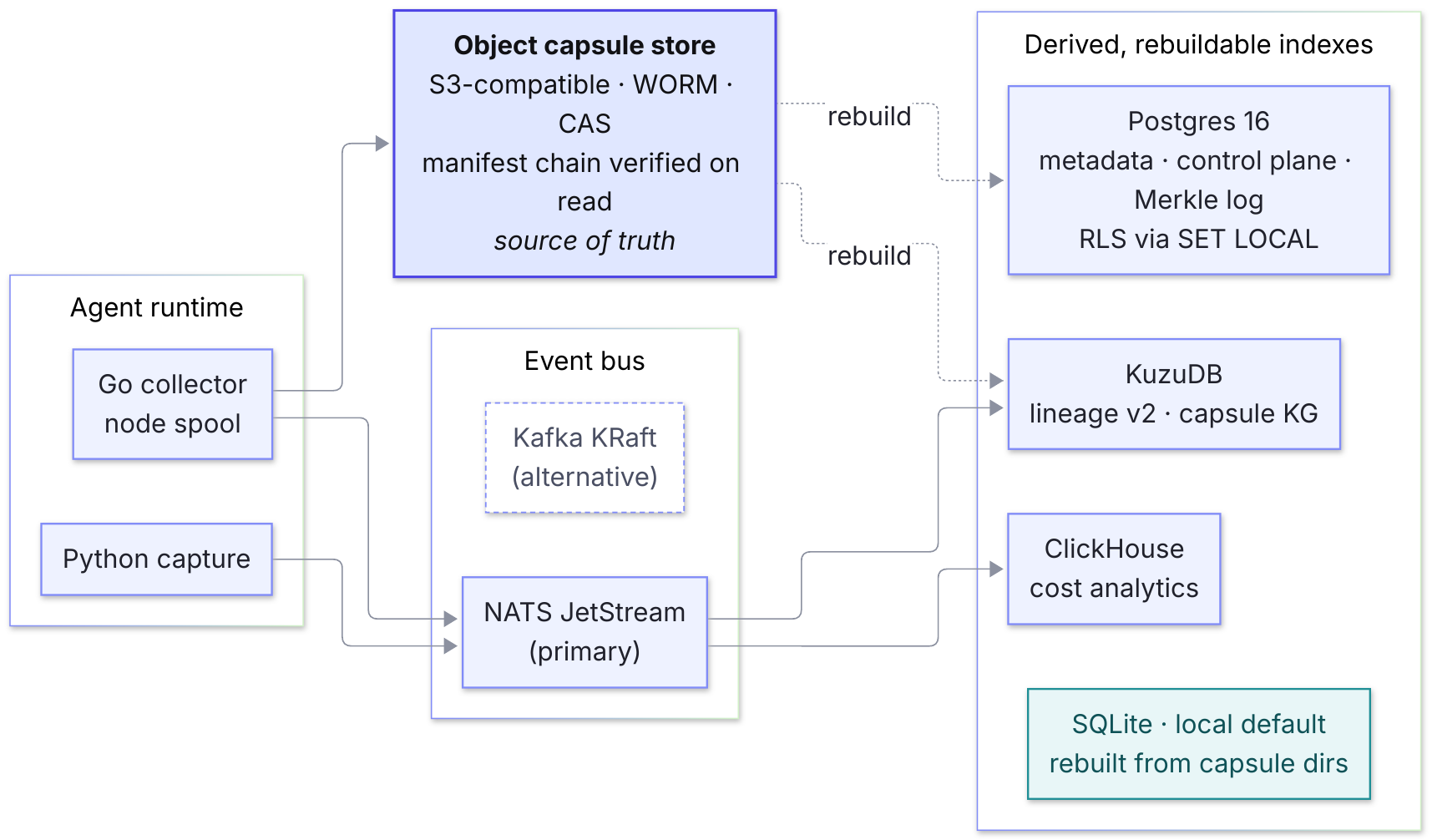}{Production storage stack. Dashed arrows are rebuilds, and
SQLite remains the zero-setup local default.}

Every backend in the stack is licensed Apache-2.0, MIT, BSD or PostgreSQL under a
default-deny dependency policy (ADR-0024). The whole stack can run self-hosted and
air-gapped. A cache tier will use Valkey rather than a source-available alternative
(\planned).

\section{Reliability contract}
The pipeline assumes distributed reality: at-least-once delivery, duplicates,
out-of-order events, missing children and untrusted clocks. If the collector is unreachable, events spool locally. If the signing
service is down, the batch is refused rather than forwarded unsigned (fail-closed). If an
index is lost, it is rebuilt from capsules. If NovaFabric itself fails, \textbf{the workload
continues}.

\chapter{Performance evidence}\label{ch:performance}

\section{One catalogue, three kinds of number}
Every published performance number lives in one file, \fpath{tests/bench/slo\_catalog.toml}
(ADR-0248). The user-facing \fpath{docs/slo.md} is generated from it, and a drift test
keeps the two identical. Each entry carries exactly one status:
\begin{itemize}
  \item \nfgated{}: a CI or nightly gate enforces it. The gate test reads its threshold
        \emph{from the catalogue}, so the published number and the enforced number cannot
        disagree.
  \item \nfmeasured{}: observed on stated hardware. It carries a
        \code{revalidate\_by} date, and once that passes, a stale claim fails the suite
        instead of fossilizing. A claim that cannot be re-measured in time is demoted to
        \nftarget{} rather than renewed.
  \item \nftarget{}: an aspiration, explicitly not a promise.
\end{itemize}
At v\nfVersion{} the catalogue holds \nfSLOs{} entries: \nfSLOGated{} gated, \nfSLOMeasured{}
measured and \nfSLOTarget{} target. The table below is generated from it.

\begin{evidence}[Why this matters]
A recorder for evidence has to hold its own claims to an evidentiary standard. In this
catalogue the published number and the enforced number are the same number. Measured
numbers expire. A claim that turned out false was withdrawn in writing, and the
original sentence was struck through rather than deleted. A reviewer can check every row below against one file and one test run.
\end{evidence}

\begin{nftable}\footnotesize
\begin{xltabular}{\linewidth}{@{}P{0.31\linewidth}P{0.16\linewidth}lL@{}}
\toprule
\thead{Catalogue entry} & \thead{Value} & \thead{Status} & \thead{Backing} \\
\midrule
\endhead
\code{capture.\allowbreak{}trivial-run.\allowbreak{}p95} & 2\,s & \nfgated & \fpath{tests/\allowbreak{}bench/\allowbreak{}test\_capture\_overhead\_gate.\allowbreak{}py} (ci:capture-overhead-gate) \\
\code{seal.\allowbreak{}seal-call.\allowbreak{}median} & 50\,ms & \nfgated & \fpath{tests/\allowbreak{}seal/\allowbreak{}test\_benchmark.\allowbreak{}py} (ci:seal-latency-gate) \\
\code{dashboard.\allowbreak{}runs-page.\allowbreak{}p95} & 100\,ms & \nfgated & \fpath{tests/\allowbreak{}bench/\allowbreak{}test\_dashboard\_scale\_gate.\allowbreak{}py} (nightly:dashboard-scale) \\
\code{dashboard.\allowbreak{}runs-search-page.\allowbreak{}p95} & 100\,ms & \nfgated & \fpath{tests/\allowbreak{}bench/\allowbreak{}test\_dashboard\_scale\_gate.\allowbreak{}py} (nightly:dashboard-scale) \\
\code{dashboard.\allowbreak{}analytics-summary.\allowbreak{}p95} & 250\,ms & \nfgated & \fpath{tests/\allowbreak{}bench/\allowbreak{}test\_dashboard\_scale\_gate.\allowbreak{}py} (nightly:dashboard-scale) \\
\code{dashboard.\allowbreak{}report-throughput.\allowbreak{}p95} & 300\,ms & \nfgated & \fpath{tests/\allowbreak{}bench/\allowbreak{}test\_dashboard\_scale\_gate.\allowbreak{}py} (nightly:dashboard-scale) \\
\code{dashboard.\allowbreak{}report-executive-summary.\allowbreak{}p95} & 300\,ms & \nfgated & \fpath{tests/\allowbreak{}bench/\allowbreak{}test\_dashboard\_scale\_gate.\allowbreak{}py} (nightly:dashboard-scale) \\
\code{dashboard.\allowbreak{}report-run-history-page.\allowbreak{}p95} & 300\,ms & \nfgated & \fpath{tests/\allowbreak{}bench/\allowbreak{}test\_dashboard\_scale\_gate.\allowbreak{}py} (nightly:dashboard-scale) \\
\code{dashboard.\allowbreak{}report-run-history-deep-keyset.\allowbreak{}p95} & 300\,ms & \nfgated & \fpath{tests/\allowbreak{}bench/\allowbreak{}test\_dashboard\_scale\_gate.\allowbreak{}py} (nightly:dashboard-scale) \\
\code{dashboard.\allowbreak{}audit-tail.\allowbreak{}p95} & 100\,ms & \nfgated & \fpath{tests/\allowbreak{}bench/\allowbreak{}test\_dashboard\_scale\_gate.\allowbreak{}py} (nightly:dashboard-scale) \\
\code{seal.\allowbreak{}seal-call.\allowbreak{}p99} & 200\,ms & \nftarget & alarm only, not a promise \\
\code{lineage.\allowbreak{}kuzu-blast-radius.\allowbreak{}p99} & 45.5\,ms & \nftarget & alarm only, not a promise \\
\code{collector.\allowbreak{}seal-batch.\allowbreak{}throughput} & 295,000\,events/s & \nftarget & alarm only, not a promise \\
\code{query.\allowbreak{}persistent-index.\allowbreak{}speedup} & 5.1x & \nfmeasured & workstation \\
\code{capture.\allowbreak{}absolute-overhead.\allowbreak{}mean} & 4.56\,s & \nfmeasured & Azure D8s\_v5 \\
\code{topology.\allowbreak{}recluster.\allowbreak{}per-node} & 17.7\,ms/node & \nfmeasured & Azure D8s\_v5 \\
\code{topology.\allowbreak{}cluster-load.\allowbreak{}p99} & 2\,ms & \nfmeasured & Azure D8s\_v5 \\
\code{policy.\allowbreak{}decision.\allowbreak{}p99} & 29\,ms & \nfmeasured & Azure D8s\_v5 \\
\code{lineage.\allowbreak{}bulk-copy.\allowbreak{}throughput} & 1,340\,edges/s & \nfmeasured & Azure D8s\_v5 \\
\code{metadata.\allowbreak{}rls-overhead.\allowbreak{}p50} & 7.4\,\% & \nfmeasured & Azure D8s\_v5 \\
\code{collector.\allowbreak{}marginal-per-event} & 0.55\,ms & \nfmeasured & Azure D8s\_v5 \\
\bottomrule
\end{xltabular}

\end{nftable}

\section{Reading the capture-overhead numbers}
Two capture figures differ by an order of magnitude, and both are correct. They measure
different things:
\begin{itemize}
  \item The \textbf{gated trivial run} (\code{python -c pass} with fast-emit hooks) had a
        median of 168.8\,ms (n\,=\,30) when it was last characterized. It is gated at
        p95 < 2\,s.
  \item The \textbf{measured full-capture mean} of 4.56\,s is a fixed cost that stays
        flat across a 61.8× range of workload sizes. Per intercepted call, the
        marginal cost is about 0.41--0.55\,ms: 3.6\,\% on an 11.5\,ms call and 0.5\,\%
        on a 102\,ms call.
\end{itemize}
Part of the fixed cost depends on \emph{which SDKs are installed} in the environment.
The default hook installer imports every present target SDK at interpreter start
purely to patch it (about 717\,ms for \code{openai} and 340\,ms for \code{mcp}, measured
with \code{-X importtime}), even when the workload never calls them. \code{--fast-emit}
patches an SDK only when the workload itself imports it. When it shipped in v0.54 it
took a compute-only workload from 2068\,ms to 464\,ms, and an \code{import openai}
workload from 2223\,ms to 1509\,ms. For long-running agent workloads the marginal cost
dominates. For very short commands the fixed cost does, and the warm capture daemon
exists for that case.

\begin{honest}[Where the numbers stop]
\begin{itemize}
  \item The ADR-0021 budget of ≤\,50\,ms for a no-op capture is a \emph{target}.
        No measured figure meets it today.
  \item The 295,000\,events/s collector batch-signing figure comes from the Phase~2
        acceptance run on a workstation, built without the race detector. CI runs the
        collector under the race detector and enforces only the latency bound. No
        re-measurement was available before its revalidation date, so since v0.107 the
        catalogue lists this figure, and the KuzuDB blast-radius figure, as \nftarget{},
        no longer as \nfmeasured{}. The project makes no cluster-scale throughput claim
        beyond these.
  \item The measurements from the 2026-08-28 cloud campaign (the full-capture mean, the
        marginal cost per call, the lineage bulk-copy rate and the topology recluster
        cost among them) cannot be renewed as they stand, because equivalent virtual
        machines are no longer available to the project. Before 2027-03-01 each is either
        re-measured on other stated hardware, as a new measurement, or demoted to
        \nftarget{}.
  \item The lineage bulk-copy path measured 1,340\,edges/s, below its own 10,000/s
        objective. Topology re-clustering crosses its 30\,s objective at about 1,700
        nodes. Both are published, not hidden.
  \item The NovaSeal log-verification target is \withdrawn{} (Chapter~\ref{ch:trust}).
        The seal p99 of 200\,ms is an alarm, not a gate: three of 55 shared-runner runs
        stalled past it.
\end{itemize}
\end{honest}

\chapter{Capability and maturity matrix}\label{ch:maturity}

The project describes itself as \textbf{beta}. The README lists as stable: local
capture, replay, diff, SQLite lineage, the trust layer, the asset registry,
policy and approval gates, and the standard eval suites. Server mode, the cluster tier
and every cohort shipped since v0.59 are experimental. The Run Capsule and Evidence
Bundle formats are not yet frozen. The condensed matrix:

\begin{nftable}
\begin{xltabular}{\linewidth}{@{}Ll@{}}
\toprule
\thead{Capability} & \thead{Status} \\
\midrule
\endhead
Wire and SDK capture, \code{record} façade, secret scan, residual rescan, redaction proof & \works \\
MCP and API capture proxies & \works \\
Runners: local, Docker, Kubernetes, Slurm, PBS, LSF; scheduler context in manifest & \exper \\
Framework adapters (\nfAdapters) & \exper \\
Warm capture daemon, OTLP ingest, sessions, computer-use capture & \exper \\
Replay: forensic, semantic, exact eligibility, mocked (fail-closed) & \works \\
Replay: intervention; non-MCP tool substitution (ADR-0306) & \exper \\
Structural diff and exit-code contract; parameter diff & \works \\
Drift detectors, replay equivalence, assurance verdicts & \exper \\
NovaSeal local profile: DSSE P-256, RFC~3161, Merkle log & \works \\
\code{identity\_trust} levels, \cmd{nova seal init} local CA & \exper \\
Cloud KMS signing, Sigstore, key ratchet, MMR, witnesses, hybrid signatures, \code{did:key}/VC & \exper \\
Post-quantum signing (ML-DSA) & \fdesign{} (registry slot only) \\
HSM / PKCS\#11 signing backend & \fdesign \\
Evidence Bundle (single and multi-capsule), admissibility status & \works \\
WORM adapters, retention sweep, legal holds (one store for CLI and API; the retention \code{crypto-shred} action is \exper) & \works \\
Maker-checker promotion and capsule attestation (signers bound by an operator trust list or pinned CA) & \works \\
Protected labels: maker-checker moves and relaxations (\cmd{nova label}) & \exper \\
Hash-chained audit log; SIEM export (JSONL, OCSF, CEF) & \works{} / \exper \\
Lineage on SQLite; OpenLineage and W3C PROV export & \works \\
Lineage on Postgres, KuzuDB, Apache AGE, JanusGraph; NATS lineage consumer & \exper \\
Capsule knowledge graph; Security Provenance KG phase~1 & \exper \\
Cross-site lineage federation & \exper{} (prototype; sovereignty review open) \\
OPA/Rego policy gates (\code{opa} sandboxed: no environment, no \code{http.send}); eval-gated promotion; a \code{promote\_direct} gate on every route into production (a rollback to a version that was live before excepted) & \works \\
Human-oversight evidence (HITL, consent), safety-case compiler, accountability spine & \exper \\
Approval taxonomy L0--L4, human-impact ledger & \fdesign \\
GDPR crypto-shredding: key lifecycle and receipts (\cmd{nova pii erase}) & \exper \\
Per-subject PII encryption at capture (\code{--pii-subject}, ADR-0314) & \exper \\
Regulatory exporters (\nfExportCmds{} commands), audit profiles (\nfAuditProfiles) & \exper \\
Cost attribution with unpriced-model honesty & \exper \\
Local dashboard (\nfDashTabs{} tabs) with route authorization & \exper \\
Server mode: OIDC, RBAC, offline tokens, SAML, SCIM, API keys & \exper \\
Go collector, object capsule store, Postgres metadata under RLS & \exper \\
Parent/child capsules: tree view, \cmd{nova run validate-distributed}, parent check on upload & \exper \\
Parent/child arrival counting, orphan placeholder and finalization (library only; not called by \cmd{nova capture}) & \exper \\
Multi-tenant dashboard serving & refused at startup until all stores are tenant-safe \\
Global query plane, cell-based fabric, multi-region log sovereignty & \fdesign \\
\bottomrule
\end{xltabular}
\end{nftable}

Three standing rules apply. No format is frozen before the v1.0 freeze. No
million-agent readiness is claimed. NovaFabric produces \emph{evidence for} compliance
workflows and never certifies compliance.

\chapter{Roadmap and future work}\label{ch:roadmap}

\section{Now: freeze the capsule format}
The single highest-leverage step is the \textbf{v1.0 capsule-format freeze}. The target
schema already exists (\fpath{schemas/run-capsule.schema.json}, \code{run-capsule-v1}),
and \cmd{nova migrate} converts v0.1.x capsules to v1.0 in a copy. The freeze is gated on sign-off
by independent design partners. The maintainer's own validation explicitly does not
count, so the gate cannot be satisfied from inside the project. Once frozen, the
format falls under the published stability policy and the RFC process, and becomes a
candidate for implementation by other tools.

\section{Planned follow-ups}
\begin{itemize}
  \item An \code{identity\_trust} level for Sigstore-signed capsules.
  \item Validating \code{--ca-bundle} chains at the RFC~3161 time rather than at
        verification time.
  \item Binding maker-checker signers to authenticated (OIDC or Sigstore) identities.
        Today they are bound to an operator trust list. The protected-stage promotion gate
        itself no longer waits on this: since v0.107 the HTTP and dashboard promote
        routes, \cmd{nova server} promotion and \cmd{nova prompt promote} go through
        \code{promote\_direct}, and so does a rollback unless its target was in production before.
  \item Capsule-scoped legal holds. Holds are registry-wide today, so any active hold
        defers every PII erasure, because one subject key covers all of that subject's
        capsules.
  \item An opt-in ``seal or abort'' capture mode.
  \item Capture-level enforcement in the SDK and wire hooks, which would close
        invariant~4. Today the \code{minimal} and \code{standard} levels do not reduce
        capture; only \code{forensic} and \code{air\_gapped} change behaviour
        (\exper).
  \item Tool substitution for LangChain/LangGraph, Pydantic AI, LlamaIndex and CrewAI
        (ADR-0306).
  \item Kubernetes \code{secretKeyRef} pass-through. Recording PBS/LSF job ids, the
        Slurm node list and the Kubernetes pod and node (ADR-0307).
  \item An external anchor writer for the adversary-anchored ledger.
\end{itemize}

\section{Longer horizon}
\begin{itemize}
  \item \textbf{Post-quantum signing.} A hybrid, crypto-agile envelope already ships
        (\exper), and ML-DSA has a reserved registry slot. Migration follows the NIST
        deprecation calendar for classical signatures (IR~8547): co-sign, then flip
        primacy, then retire.
  \item \textbf{Federation.} Cross-organization provenance with cryptographic trust
        between clusters that share no infrastructure. Today there is a stateless
        fan-out query coordinator; the protocol is \fdesign.
  \item \textbf{Workload identity.} SPIFFE/SPIRE and broader DID/VC identity beyond
        today's \code{did:key} slice.
  \item \textbf{Cell-based fabric.} Operationally independent cells under regional
        and global control planes, with placement driven by residency, cost and carbon.
\end{itemize}

\section{Not planned}
Real-time monitoring and alerting, a hosted SaaS, and a prompt playground are out of
scope by decision. They belong to adjacent tools that NovaFabric integrates with.

\chapter{Engineering practice, standards and security}\label{ch:practice}

\section{How it is built}
\begin{itemize}
  \item \textbf{Quality gates.} pytest with a 90\,\% coverage floor (\code{--cov-fail-under=90}
        in CI), mypy in strict mode, and ruff. The suite has \nfTestFuncs{} test
        functions in \nfTestFiles{} files, run in load-aware parallel tiers. The
        collector CI runs \code{go vet}, the race detector, coverage and
        \code{govulncheck}. A dedicated security gate covers the metadata store.
  \item \textbf{\nfWorkflows{} CI workflows.} Unit and container tiers, nightly scale
        gates, licence policy, pip and npm audit, MCP conformance, OpenSSF Scorecard,
        and signed releases to PyPI, GHCR and an OCI Helm registry.
  \item \textbf{Decision discipline.} \nfADRs{} ADRs, numbered up to ADR-0\nfADRMax{} and kept in the
        maintainers' private design repository, record every consequential decision. A redesign must be justified by a blocked
        capability, a failing test or benchmark, an insufficient simpler migration, an
        ADR, and preserved compatibility.
  \item \textbf{Documentation as a tested artefact.} Guard tests fail when documented
        counts drift from the code. This covers the CLI reference, the SLO page, the
        route table and the differentiation table. Every doc mentioning a future feature
        must carry one of the four maturity labels.
  \item \textbf{Licence policy (ADR-0024).} Tier~A (Apache-2.0, MIT, BSD, PostgreSQL,
        ISC, PSF, CC0) is allowed. Tier~B (LGPL dynamic, MPL-2.0, EPL-2.0) needs review.
        GPL, AGPL, SSPL, BSL, Elastic and source-available licences are rejected unless
        explicitly waived.
\end{itemize}

\section{Standards alignment}
\begin{nftable}
\begin{tabularx}{\linewidth}{@{}P{0.34\linewidth}Ll@{}}
\toprule
\thead{Standard} & \thead{Where it is used} & \thead{Level} \\
\midrule
in-toto Statement v1, DSSE & evidence bundles, seals, promotion attestations & \works \\
RFC~3161, RFC~8785, RFC~6962-style proofs & timestamps, canonical digests, Merkle proofs & \works \\
JSON Schema 2020-12 & capsule, envelope and bundle schemas & \works \\
OpenTelemetry GenAI semconv & model and tool call payloads; normative envelope mapping & \works \\
OTLP trace and log ingest & capsules built from external OTel telemetry & \exper \\
OpenLineage, W3C PROV-JSON / PROV-N & lineage emission and export & \works \\
W3C PROV-O, SHACL, openCypher & semantic knowledge-graph surfaces & \exper \\
SLSA provenance & release artefacts & \works \\
CloudEvents~1.0 & alternative envelope mapping (normative in Envelope v1) & \exper \\
CycloneDX~1.7 AI/ML, C2PA, RO-Crate, BagIt & AIBOM, provenance marking, archival & \exper \\
MCP (incl.\ server cards) & tool-call capture, proxy, scanner, conformance & \exper \\
OIDC, RFC~8628, SAML~2.0, SCIM~2.0 & server-mode identity & \exper \\
EU AI Act, NIST AI RMF, ISO/IEC~42001, GDPR, NIS2, HIPAA & exporters and audit profiles & \exper \\
\bottomrule
\end{tabularx}
\end{nftable}

\section{Security posture}
\begin{itemize}
  \item \textbf{No network surface by default.} Server mode, webhooks, SCIM, SAML, OTLP
        ingest and remote runners are all opt-in. Settings that disable a safety
        control are explicit environment variables, and each one is listed.
  \item \textbf{Secrets are meant to stay in the redaction boundary.} Scan, redact, residual rescan,
        redaction proof; the scanner knows patterns, not secrets, so this is an objective with
        stated limits (Chapter~\ref{ch:capture}), not a guarantee. Off-host runners never forward the submitting shell: Docker and
        Kubernetes pass an allowlist (ADR-0270), SLURM and LSF submit from a minimal
        environment (ADR-0311), and no runner forwards a NovaFabric credential. Anything
        else a job needs is passed explicitly with \code{extra\_env}. Since v0.107 such
        a value stays off the command lines the runner starts: Slurm and LSF send the
        job script on standard input, Docker passes \code{-e NAME} with the value in the
        client's environment (except a name that steers the client itself, such as
        \code{PATH} or \code{DOCKER\_*}), and Slurm, LSF and PBS quote with
        \code{shlex.quote}.
  \item \textbf{The capsule is untrusted once the workload has run.} A sweep after the
        workload removes every symbolic link, FIFO, socket and device at any depth
        without following it, records the tamper in the signed manifest
        (\code{capture\_tampered}) and pins the capsule directory. \cmd{nova verify}
        and \cmd{nova validate} fail such a capsule, and exporters, \cmd{nova serve}
        routes and backups never follow a link. A capture hook that fails never reaches
        the workload (fail-open, counted in \code{capture-health.json}), and schema
        validation never retrieves a remote \code{\$ref}, even at \code{air\_gapped}.
  \item \textbf{One verifier per artefact.} \cmd{nova verify}, the trust radar, the Seal
        tab and \code{POST /api/runs/\{id\}/verify} share one capsule verifier, and
        \cmd{nova verify <bundle>.zip} and \code{POST /api/evidence/\{id\}/verify} share
        one bundle verifier, so a tampered capsule cannot read as attested on one
        surface and fail on another. A bundle's members are all signed, and signer trust
        is pinned with \code{NOVAFABRIC\_EVIDENCE\_TRUSTED\_KEYS}.
  \item \textbf{No root containers, no privileged Kubernetes.} The shipped image's user
        id is pinned to the chart's by a guard test. The chart is secure by default
        (ADR-0230).
  \item \textbf{Supply chain.} Keyless-cosign-signed images with SBOM and provenance,
        Trivy release scans, pip/npm audit gates whose waivers expire, and OIDC Trusted
        Publishing to PyPI.
  \item \textbf{Disclosure.} Private reporting with acknowledgement within five business
        days and severity-based fix targets (\fpath{SECURITY.md}).
\end{itemize}

\begin{honest}[Stated security boundaries]
No FIPS validation is claimed, and Ed25519-only surfaces have no FIPS-approved path.
ML-DSA is not implemented. The SAML implementation is still awaiting a dedicated
security-architecture review. Secret scanning is pattern-based (Chapter~\ref{ch:capture}).
Hard links inside a capsule are not detected by the exporters. An \code{extra\_env} value
given as \code{--runner-option extra\_env=K=V} is still in \cmd{nova capture}'s own
argument list. A same-host reverse proxy that adds no forwarding header looks local to the
cluster collector, which is why it needs a token whenever anything proxies to it. Audit
records written before v0.107 still hold the first eight characters of the \cmd{nova serve}
token, because the hash-chained log is not rewritten.
\end{honest}

\chapter{Conclusion}\label{ch:conclusion}

AI systems became consequential before they became accountable. NovaFabric's working
assumption is that the unit of accountability can be a \textbf{sealed, replayable evidence
capsule}. Whether that assumption holds beyond one project depends on independent
verifiers, design partners and adopters; none of that evidence is in this report.

Between Edition~1 and this edition the important change is not the larger command
count. It is that the engines now \emph{refuse} rather than approximate:
\begin{itemize}
  \item mocked replay fails closed on any divergence;
  \item the diff gate distinguishes ``changed'' from ``could not compare'';
  \item the dashboard denies any route it has not classified;
  \item multi-tenant serving refuses to start while a store is unsafe;
  \item a cost aggregate refuses rather than print a misleading zero;
  \item an erasure command that could not erase now says so, and \cmd{nova pii erase}
        reaches PII encrypted at capture (\exper, opt-in per subject) and defers while any
        legal hold is active;
  \item a signature, an Evidence Bundle or a maker-checker approval no longer passes on
        a claim it does not cover, and the dashboard's trust views run the same verifier
        as \cmd{nova verify};
  \item a link or special file planted in a capsule by the workload is removed and
        recorded, never followed, and fails verification;
  \item a performance target that could not be met was withdrawn in writing.
\end{itemize}
A system whose job is to make \emph{other} systems tell the truth needs exactly this
property.

The remaining work is clear and written down. Freeze the capsule format with
independent design partners. Bring prompt capture under the capture-level policy.
Re-measure the numbers that are due
for revalidation. Grow the cluster tier from
experimental to proven on real HPC hardware. Every item has an ADR, a roadmap entry and
a maturity label.

\vspace{6mm}
{\sffamily\small\color{nfSlate}
Source, documentation and releases: \textbf{\color{nfViolet}github.com/MSKazemi/novafabric}\\
Author: Mohsen Seyedkazemi Ardebili \textperiodcentered{} \textbf{\color{nfViolet}mskazemi.github.io}}

\appendix
\chapter{CLI command surface}\label{app:cli}

All \nfTopCmds{} top-level \cmd{nova} commands at v\nfVersion, generated from the Typer
application by \code{novafabric.\allowbreak cli.\allowbreak introspect}. A superscript gives the number of
runnable sub-command paths under a group; a name without one is a single command. In
total there are \nfCmdPaths{} runnable paths.

\vspace{2mm}
{\raggedright\linespread{1.5}\selectfont
\cliitem{a2a\mbox{-}card}{2}
\cliitem{a2a\mbox{-}objects}{2}
\cliitem{aibom}{3}
\cliitem{annotate}{9}
\cliitem{api\mbox{-}proxy}{0}
\cliitem{approve}{0}
\cliitem{asset}{1}
\cliitem{assure}{0}
\cliitem{assure\mbox{-}alarm}{1}
\cliitem{assure\mbox{-}baseline}{2}
\cliitem{assure\mbox{-}canary}{1}
\cliitem{assure\mbox{-}case}{0}
\cliitem{assure\mbox{-}coverage}{0}
\cliitem{assure\mbox{-}impact}{1}
\cliitem{assure\mbox{-}run}{2}
\cliitem{audit}{5}
\cliitem{audit\mbox{-}log}{2}
\cliitem{backup}{2}
\cliitem{capsule}{1}
\cliitem{capture}{0}
\cliitem{capture\mbox{-}ui}{2}
\cliitem{classify}{3}
\cliitem{collector}{1}
\cliitem{comment}{3}
\cliitem{consent}{6}
\cliitem{cost}{6}
\cliitem{daemon}{3}
\cliitem{dashboard}{5}
\cliitem{dataset}{1}
\cliitem{db}{2}
\cliitem{diagnose}{0}
\cliitem{diff}{0}
\cliitem{doctor}{0}
\cliitem{drift}{5}
\cliitem{dsar}{2}
\cliitem{embodied}{6}
\cliitem{energy}{4}
\cliitem{erasure}{2}
\cliitem{euaiact}{2}
\cliitem{eval}{20}
\cliitem{events}{2}
\cliitem{evidence}{5}
\cliitem{experiment}{4}
\cliitem{export}{0}
\cliitem{export\mbox{-}accessibility\mbox{-}claim}{0}
\cliitem{export\mbox{-}adverse\mbox{-}action}{0}
\cliitem{export\mbox{-}aibom}{0}
\cliitem{export\mbox{-}annex\mbox{-}iv}{0}
\cliitem{export\mbox{-}blob}{0}
\cliitem{export\mbox{-}c2pa}{0}
\cliitem{export\mbox{-}cat}{0}
\cliitem{export\mbox{-}citizen\mbox{-}explanation}{0}
\cliitem{export\mbox{-}compliance}{4}
\cliitem{export\mbox{-}control\mbox{-}attestation}{0}
\cliitem{export\mbox{-}election\mbox{-}disclosure}{0}
\cliitem{export\mbox{-}evidence}{0}
\cliitem{export\mbox{-}examiner}{3}
\cliitem{export\mbox{-}foia}{0}
\cliitem{export\mbox{-}hipaa\mbox{-}proof}{0}
\cliitem{export\mbox{-}model\mbox{-}independence}{0}
\cliitem{export\mbox{-}model\mbox{-}risk}{0}
\cliitem{export\mbox{-}nis2}{0}
\cliitem{export\mbox{-}nist\mbox{-}rmf}{0}
\cliitem{export\mbox{-}part11}{0}
\cliitem{export\mbox{-}public\mbox{-}annex\mbox{-}viii}{0}
\cliitem{export\mbox{-}public\mbox{-}disclosure}{0}
\cliitem{export\mbox{-}public\mbox{-}incident}{0}
\cliitem{export\mbox{-}rai\mbox{-}scorecard}{0}
\cliitem{export\mbox{-}retention}{0}
\cliitem{export\mbox{-}rocrate}{0}
\cliitem{export\mbox{-}rocrate\mbox{-}science}{0}
\cliitem{export\mbox{-}ropa}{0}
\cliitem{export\mbox{-}system\mbox{-}card}{0}
\cliitem{export\mbox{-}transparency\mbox{-}register}{0}
\cliitem{export\mbox{-}whistleblower}{0}
\cliitem{forensics}{1}
\cliitem{graph}{1}
\cliitem{hitl}{7}
\cliitem{hold}{3}
\cliitem{import}{0}
\cliitem{incident}{4}
\cliitem{ingest\mbox{-}capsule}{0}
\cliitem{init}{0}
\cliitem{insights}{0}
\cliitem{inspect}{0}
\cliitem{insurance}{3}
\cliitem{kg}{16}
\cliitem{label}{8}
\cliitem{ledger}{3}
\cliitem{lineage}{11}
\cliitem{lineage\mbox{-}store}{2}
\cliitem{list}{0}
\cliitem{login}{0}
\cliitem{logout}{0}
\cliitem{mcp}{5}
\cliitem{mcp\mbox{-}proxy}{0}
\cliitem{media}{1}
\cliitem{memory}{2}
\cliitem{memstore}{3}
\cliitem{merkle\mbox{-}tree}{0}
\cliitem{migrate}{0}
\cliitem{migrate\mbox{-}format}{0}
\cliitem{migrate\mbox{-}schema}{0}
\cliitem{migrate\mbox{-}to\mbox{-}postgres}{0}
\cliitem{new\mbox{-}run\mbox{-}id}{0}
\cliitem{passport}{2}
\cliitem{pii}{2}
\cliitem{policy}{6}
\cliitem{preservation}{4}
\cliitem{pricing}{3}
\cliitem{promote}{4}
\cliitem{prompt}{8}
\cliitem{provenance}{5}
\cliitem{query}{0}
\cliitem{rebuild\mbox{-}metadata\mbox{-}db}{0}
\cliitem{redact}{0}
\cliitem{redaction\mbox{-}xray}{0}
\cliitem{register}{0}
\cliitem{replay}{0}
\cliitem{replay\mbox{-}equivalence}{2}
\cliitem{report}{0}
\cliitem{restore}{0}
\cliitem{retention}{4}
\cliitem{rollback}{0}
\cliitem{run}{4}
\cliitem{safety}{3}
\cliitem{safety\mbox{-}case}{3}
\cliitem{scan\mbox{-}secrets}{0}
\cliitem{schema}{1}
\cliitem{science}{5}
\cliitem{score}{1}
\cliitem{seal}{10}
\cliitem{search}{0}
\cliitem{serve}{0}
\cliitem{server}{15}
\cliitem{session}{9}
\cliitem{settlement}{1}
\cliitem{storage}{2}
\cliitem{subject\mbox{-}proof}{0}
\cliitem{suggest\mbox{-}register}{0}
\cliitem{support\mbox{-}bundle}{0}
\cliitem{toolschema}{4}
\cliitem{trend}{0}
\cliitem{trust\mbox{-}path}{2}
\cliitem{trust\mbox{-}radar}{0}
\cliitem{unregister}{0}
\cliitem{validate}{0}
\cliitem{verify}{0}
\cliitem{verify\mbox{-}envelope}{0}
\cliitem{view}{5}

\par}

\section*{Selected groups}
\begin{nftable}
\begin{tabularx}{\linewidth}{@{}lL@{}}
\toprule
\thead{Group} & \thead{Sub-commands} \\
\midrule
\cmd{nova seal} & init, sign, verify, propose, approve, bypass, log verify, ratchet init / rotate / status \\
\cmd{nova lineage} & provenance, blast-radius, replay-chain, root-cause, time-travel, import, consume, metrics, export-prov, export-graph, emit-openlineage \\
\cmd{nova kg} & init, build, build-provenance, ingest, query, status, audit, detect, attack-path, blast-radius, alias, entity-queue \\
\cmd{nova policy} & test, explain, sign, list, capture-level get / set \\
\cmd{nova promote} & direct, propose, approve, keygen \\
\cmd{nova label} & set, get, list, history, protect, propose-move, approve-move, status \\
\cmd{nova hold} & create, list, release \\
\cmd{nova audit} & map, report, verify, bundle, coverage \\
\cmd{nova eval} & run, agent, compare, list, offline, cost, contamination-check, card, score, export-inspect, import-inspect \\
\cmd{nova server} & start, issue-token, revoke-token, assign-role, revoke-role, api-key, saml-metadata, scim-map-group, list-scim-events, usage, flush-jwks-cache \\
\cmd{nova session} & new, add, list, show, replay, export, import, reindex, verify-bundle \\
\cmd{nova evidence} & attest-replay, bind, bind-custody, check-admissibility, completeness \\
\cmd{nova retention} & plan, apply, status, explain \\
\cmd{nova run} & new-run-id, show, lineage, validate-distributed \\
\bottomrule
\end{tabularx}
\end{nftable}

\chapter{What changed since Edition 1}\label{app:changes}

Edition~1 described v0.45.1 (June 2026). The most significant additions since then,
taken from \fpath{CHANGELOG.md}:

\begin{nftable}
\begin{xltabular}{\linewidth}{@{}lL@{}}
\toprule
\thead{Version} & \thead{Change} \\
\midrule
\endhead
v0.49 & CloudEvents envelope interop; span taxonomy grows to 33 event types; \cmd{nova diagnose} failure attribution; significance gate on promotion \\
v0.50 & intervention (counterfactual) replay; forward-secure key ratchet; Merkle consistency proofs; column-level lineage \\
v0.52 & warm capture daemon (\cmd{nova daemon}) \\
v0.55--0.56 & the accountability spine: energy receipts, adversary-anchored ledger, replay attestation, safety-case compiler, admissibility binding \\
v0.59 & session capsules and session replay; multi-modal media capture; agent execution graph; OTLP trace ingest; first SAML/SCIM slices; local pricing catalogue \\
v0.61 & offline drift detectors and the first assurance-loop slices \\
v0.64 & additive capsule \code{facets} container; evidence-grade backup and restore of every local store \\
v0.68--0.70 & Postgres, Apache AGE and JanusGraph lineage backends become real \\
v0.71--0.72 & AWS, GCP and Azure KMS key-wrapping (envelope encryption) \\
v0.75--0.91 & witness co-signing, delegation chains, MMR, \code{did:key}/VC, hybrid signatures, jurisdiction site-seals, pinned x509 identity \\
v0.80--0.89 & EU AI Act (Art.~12/50/72, GPAI Art.~53), ISO~42001/42005 and NIST GenAI exporters; \cmd{nova export-compliance} \\
v0.101 & enterprise programme ADRs 0240--0249: air-gap bundle, residency routing, step-up auth, per-tenant KEKs, writer lease with fencing, listener TLS, durable jobs, SLO catalogue \\
v0.102 & \cmd{nova replay-equivalence}, \cmd{nova assure-*}, drift collection, \code{diff --graph-shape/--media}, evidence cart, offline TSA chain and CRL checks, LlamaIndex / Pydantic AI / Haystack adapters, isolated concurrent captures, secure-by-default chart \\
v0.103--0.104 & governed capsule delete; OTLP logs ingest; strict capture mode; secret scanner covers every ADR-0009 file \\
v0.105 & fail-closed mocked replay with MCP, async, streaming, Responses API and recorded errors (ADR-0300/0304); SDK wire records as transport (ADR-0305); \code{record.tool} and tool overrides (ADR-0306); diff exit-code contract and parameter diff (ADR-0303); residual secret rescan; scheduler context in manifest (ADR-0307); \cmd{nova seal init} and \code{identity\_trust} (ADR-0301) \\
v0.106 & signing intent inside the signature (ADR-0309); capsule Merkle root commits to file names (ADR-0310); SLURM/LSF jobs submit from a minimal environment (ADR-0311); registry signer trust list and gated direct promotion (ADR-0312); size-committing log roots (ADR-0313); per-subject PII encryption at capture (ADR-0314); maker-checker trust root (ADR-0315); \cmd{nova verify} checks Evidence Bundle signatures and subjects; exact attestation refuses ineligible runs; ADR-0306 slices 3--4 (Google ADK, OpenAI Agents); ADR-0308 eval images \\
v0.107 & two P0 fixes: caller-supplied Rego on \cmd{nova serve} needs \code{admin} and every \code{opa} runs with no inherited environment and without \code{opa.runtime}/\code{http.send}; \code{POST /v0/evidence} no longer takes a client path. Capsule treated as untrusted after the workload (post-run sweep of links, FIFOs and devices; no exporter or \cmd{nova serve} route follows a link; \cmd{nova export-evidence} refuses a capsule containing a symlink); capture never fetches a schema \code{\$ref}, and capture hooks are fail-open. One verifier per artefact behind \cmd{nova verify}, the trust radar, the Seal tab and the verify routes; every Evidence Bundle member signed, signer trust pinned by \code{NOVAFABRIC\_EVIDENCE\_TRUSTED\_KEYS}; \code{timestamp\_trust} separates an anchored from an unanchored RFC~3161 token; \code{--strict} for \cmd{nova verify}. Every route into production gated (registration accepts only \code{development}; HTTP, \cmd{nova server} and prompt promotion go through \code{promote\_direct}, and so does a rollback unless the target was in production before); protected-label maker-checker, and relaxing a protection, bound to the signer trust list; one legal-hold store that blocks erasure, deletion and the retention sweep; \cmd{nova serve} path allowlist; authenticated, tenant-scoped collector. Behaviour changes in many places; the CHANGELOG lists them under ``Changed, BREAKING'' \\
\bottomrule
\end{xltabular}
\end{nftable}

\section*{Corrections to Edition 1}
Edition~1 was accurate for its time, but five statements no longer hold:
\begin{itemize}
  \item \textbf{Capture levels.} Edition~1 listed seven capture levels. The policy
        enumeration in code has four: \code{minimal}, \code{standard}, \code{forensic}
        and \code{air\_gapped}. Only the last two change what is recorded today; the
        metadata-only allowlist of \code{minimal} is applied by no capture path.
  \item \textbf{Merkle-log verification.} The ``sampled fast path, p99 < 200\,ms at 1M
        entries'' claim is withdrawn (ADR-0268). Full recomputation is the guarantee.
  \item \textbf{Parent/child capsules.} The orphan timeout is 24\,h, not 300\,s. The
        roles are \code{PARENT}, \code{CHILD} and \code{STANDALONE}, with a separate
        \code{DRIVER} / \code{WORKER} distribution role; there is no \code{COORDINATOR}.
  \item \textbf{Seal latency.} Edition~1 listed ``NovaSeal seal latency p99 < 200\,ms''
        as CI-gated. Today the gate is the \emph{median} (< 50\,ms). The p99 is an alarm
        only, because shared CI runners occasionally stall past it.
  \item \textbf{Approval taxonomy.} The L0--L4 approval taxonomy and the human-impact
        ledger are future design, not experimental code.
\end{itemize}

\chapter{How this document was verified}\label{app:method}

\section*{Generated numbers}
Every count in this document is a LaTeX macro written by \fpath{verify\_facts.py}. The
script runs before each build against the public repository at the release tag, except the
ADR row, which reads the maintainers' private decision records. A probe that returns
nothing stops the build.

\begin{nftable}
\begin{tabularx}{\linewidth}{@{}P{0.25\linewidth}Lr@{}}
\toprule
\thead{Fact} & \thead{Probe} & \thead{Value} \\
\midrule
CLI surface & \code{cli.introspect.top\_level\_command\_names()} / \code{command\_paths()} & \nfTopCmds{} / \nfCmdPaths \\
Exporters & command paths starting \code{export-} & \nfExportCmds \\
Subpackages, size & directories and \code{*.py} line count under \fpath{src/novafabric} & \nfPackages{} / \nfPyKLOC K \\
Tests & \code{test\_*.py} files; \code{def test\_} functions under \fpath{tests/} & \nfTestFiles{} / \nfTestFuncs \\
ADRs (private design records) & \fpath{design/adr/NNNN-*.md}, highest number & \nfADRs{} / 0\nfADRMax \\
Dashboard & \code{NAV\_GROUPS} in \fpath{ui/dashboard/.../Sidebar.tsx} & \nfDashTabs{} / \nfDashGroups \\
HTTP routes & \code{serve.introspect.route\_methods()} on both app factories & \nfServeRoutes{} / \nfServerRoutes \\
Authorization table & \code{len(serve.authz.ROUTE\_SCOPES)} & \nfAuthzRoutes \\
Runners, adapters, audit profiles & runner registry; adapter modules; profile YAML files & \nfRunners{} / \nfAdapters{} / \nfAuditProfiles \\
SLO catalogue & \fpath{tests/bench/slo\_catalog.toml} by status & \nfSLOGated{}/\nfSLOMeasured{}/\nfSLOTarget \\
CI workflows & \fpath{.github/workflows/*.yml} & \nfWorkflows \\
Differentiation & \fpath{scripts/verify\_differentiation\_table.py --format json} & \nfDiffPass{}/\nfDiffClaims \\
\bottomrule
\end{tabularx}
\end{nftable}

\paragraph{Reproducing the counts.} The script \fpath{reproduce\_counts.py}, shipped with this
report's source, runs the same probes against a clone of the public repository
(\texttt{git checkout v\nfVersion}) and prints one line per fact; on that tag it prints the
values in the table. Every count can be re-derived this way except the ADR row: the decision
records live in a private repository, so that figure is the maintainers' statement and not an
independently checkable one. The build script \fpath{verify\_facts.py} differs from the public
script only in writing the LaTeX macros, the command appendix and the SLO table.

\section*{Checked by hand against the code}
Qualitative statements were read against the implementation rather than taken from the
documentation. Examples:
\begin{itemize}
  \item replay modes and the re-execution split (\fpath{replay/\_flags.py},
        \fpath{replay/\_replayability.py});
  \item the safety-ladder flags; the capture-level enumeration;
  \item the parent/child roles, states and 24-hour orphan timeout
        (\fpath{capsule/schema.py}, \fpath{capsule/orphan.py});
  \item the lineage edge vocabulary;
  \item the order of the separation-of-duties check and the policy gate in promotion
        (\fpath{registry/service.py});
  \item the five attestation checks and their exit codes (\fpath{promote/verifier.py});
  \item the OPA fallback (\fpath{policy/\_\_init\_\_.py}); S3 Object Lock mode;
  \item the signing-log root versions (\fpath{trust/novaseal/merkle.py}), re-executed:
        under the legacy v1 root the leaf lists $[a,b,c]$ and $[a,b,c,c]$ share a root, so
        a v1 root is verified only with its tree size, and under the v2 root, which every
        new log uses, they differ;
  \item the \cmd{nova erasure} correction (both subcommands are marked not implemented;
        erasure is \cmd{nova pii erase});
  \item the six v0.106 security fixes (signing intent, capsule-root renames, Evidence
        Bundle verification, bypass trust, exact attestation, SLURM/LSF environment),
        each attacked end to end against the code and found blocked.
\end{itemize}
Every \cmd{nova} command named in the text was checked against the generated command
list. As a final step, a separate AI-assisted review pass was given only the finished
text and the source tree and cross-checked about 85 claims against the code. Each
discrepancy it raised was re-verified against the code and corrected before release, among them the strength of the maker-checker identity binding, the wiring of
crypto-shredding, the replay-flag policy gate, parent finalization semantics and collector
deduplication.

\section*{Update to v0.107.0}
The second edition was first validated against v0.106.1. For v0.107.0 every chapter was read again
against the code at the \texttt{v0.107.0} tag and against the CHANGELOG entry for that
release; this was a claim-by-claim comparison of the text with the code, not a new
adversarial review. Earlier known gaps were re-checked individually in the code: the
registry maker-checker verifies both signatures against the signer trust list and compares
bound identities (\fpath{registry/service.py}, \fpath{registry/signers.py}); the promote
verifier requires its signers to chain to a pinned trust root (\fpath{promote/verifier.py});
per-subject PII encryption at capture is wired and erasure reaches it, but remains
experimental and is deferred while any legal hold is active; the replay policy gate covers
every write-permitting ladder flag (\fpath{replay/\_engine.py}); parent finalization
still counts arrivals, not child status (\fpath{capsule/writer.py}); and the MCP hook still
records full arguments and results at every capture level, because mocked replay serves
them. The generated counts were read from the \texttt{v0.107.0} release tree, not from a working
tree that may have moved on.

\section*{Authorship}
AI assistants were used in preparing this document and in the review pass described
above.

\section*{Diagrams}
All sixteen figures are Mermaid sources in \fpath{figures/*.mmd}, rendered with the
\code{mska-flowchart} toolchain and one shared theme (\fpath{figures/mermaid-config.json}).
Each was inspected as a raster image before being printed to a vector PDF. Layouts were
reworked until every label prints at 7\,pt or more. Three figures were corrected against the code: maker-checker (the policy gate
runs after the separation-of-duties check, not before), the cluster planes (deduplication
happens downstream of the collector) and parent finalization (arrival counts, not child
status).

\section*{Typography}
Source Serif~4 (text), Inter and Inter Display (headings, diagrams) and JetBrains Mono (code),
all SIL Open Font Licence, bundled in \fpath{fonts/}; typeset with XeLaTeX.

\renewcommand{\bibname}{References}

\end{document}